\documentclass[%
 reprint,
superscriptaddress,
 amsmath,amssymb,
 aps,
]{revtex4-2}

\usepackage{physics}
\usepackage{graphicx}% Include figure files
\usepackage{dcolumn}% Align table columns on decimal point
\usepackage{bm}% bold math
\usepackage{color,sidecap,colortbl}
\usepackage[dvipsnames]{xcolor}
\newcommand{\kbar}{\mathchar'26\mkern-9mu k}
\usepackage{upgreek}
\begin{document}

% \title{Prethermal Dynamical Localization and the Emergence of Chaos in a Kicked Interacting Quantum Gas}

\title{Interaction-driven breakdown of dynamical localization in a kicked quantum gas}

\author{Alec Cao}
\author{Roshan Sajjad}
\author{Hector Mas}
\author{Ethan Q. Simmons}
\author{Jeremy L. Tanlimco}
\author{Eber Nolasco-Martinez}
\author{Toshihiko Shimasaki}
\author{H. Esat Kondakci}
\affiliation{Department of Physics, University of California, Santa Barbara, California 93106, USA}

\author{Victor Galitski}
\affiliation{Joint Quantum Institute and Department of Physics, University of Maryland, College Park, MD 20742, USA}

\author{David M. Weld}
\affiliation{Department of Physics, University of California, Santa Barbara, California 93106, USA}

%\begin{abstract}
% While ergodicity is a fundamental postulate of statistical mechanics and implies that driven interacting systems inevitably heat, ergodic dynamics can be disrupted by quantum interference.

% The interplay between interactions, localization, and dynamics is a topic at the forefront of the modern understanding of condensed matter. Despite a quarter-century of experimental studies, the effect of many-body interactions on dynamically localized quantum states has remained unexplored. We report the experimental realization of a tunably-interacting kicked quantum rotor ensemble using a Bose-Einstein condensate in a pulsed optical lattice. We observe signatures of a prethermal localized plateau, followed for interacting samples by interaction-induced anomalous diffusion with an exponent near one half. Echo-type time reversal experiments establish the role of interactions in destroying reversibility.
% These results elucidate the dynamical transition to many-body quantum chaos, and explore and delimit possibilities for globally protecting quantum information in interacting driven  systems.
% \end{abstract}

\maketitle
\textbf{Quantum interference can terminate energy growth in a continually kicked system, via a single-particle ergodicity-breaking mechanism known as dynamical localization. The effect of many-body interactions on dynamically localized states, while important to a fundamental understanding of quantum decoherence, has remained unexplored despite a quarter-century of experimental studies. We report the experimental realization of a tunably-interacting kicked quantum rotor ensemble using a Bose-Einstein condensate in a pulsed optical lattice. We observe signatures of a prethermal localized plateau, followed for interacting samples by interaction-induced anomalous diffusion with an exponent near one half. Echo-type time reversal experiments establish the role of interactions in destroying reversibility. These results quantitatively elucidate the dynamical transition to many-body quantum chaos, advance our understanding of quantum anomalous diffusion, and delimit some possibilities for protecting quantum information in interacting driven systems.}

Ergodicity breaking in quantum matter and relaxation dynamics of thermalizing phases are two aspects of a central question of non-equilibrium many-body physics: how and when do isolated quantum systems thermalize? A growing body of theoretical and experimental work suggests that ergodicity can be avoided or hindered by a variety of mechanisms, including many-body localization~\cite{mblreview_huse,mblreview_abanin}, quantum many-body scarring~\cite{bernien_scars_2017,scarsreview_2021}, and prethermalization~\cite{prethermal_weld,prethermal_bloch,prethermal_cappellaro,prethermal_schmiedmayer,ueda_prethermalreview_2020}. %Often accompanying this modified thermalization is the signature of anomalous diffusion, a process classically used to describe dynamics in crowded systems (ie. protein motion in cells) \cite{anomalousprotein}. 
Even without ergodicity breaking, the expected emergence of quantum chaos upon the addition of interactions to driven systems is not well understood.
For example, one ubiquitous but incompletely-understood feature of the interface between localized and ergodic regimes is anomalous diffusion~\cite{mblsubdiffusion_demler,prethermal_subjoule_ON_knap}, which can potentially serve as an indicator of entanglement spreading~\cite{anomalousspin}. %Beyond total ergodicity breaking, prethermalization and anomalous diffusion are features of the poorly-understood interface between chaotic and localized regimes, and 
A general predictive understanding of such phenomena remains an open challenge to theory and experiment.

\begin{figure}[t!]
    \centering
    \renewcommand{\figurename}{Fig.}
    \includegraphics[scale = 1]{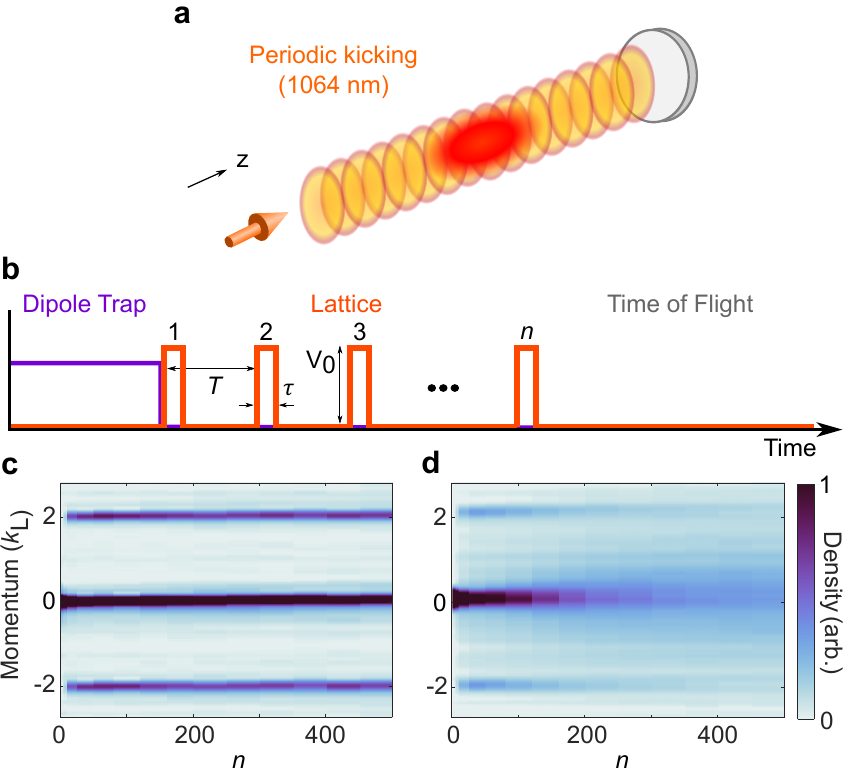}
    \caption{\textbf{Experimentally realizing an interacting quantum kicked rotor.} (\textbf{a}) Schematic of BEC in single pulsed optical lattice. (\textbf{b}) Experimental sequence. After setting the scattering length the trap is removed and kicking is applied with period $T$, pulse width $\tau$, and amplitude $V_0$ for $n$ cycles. The atoms are imaged after a time-of-flight expansion~\cite{supp}. (\textbf{c-d}) Measured axial momentum distribution versus kick number $n$ for noninteracting (c) and interacting (d) samples, revealing collisional momentum redistribution. 
    }
    \label{fig:setup}
\end{figure}

\begin{figure*}[t]
    \centering
    \renewcommand{\figurename}{Fig.}
    \includegraphics[scale=1]{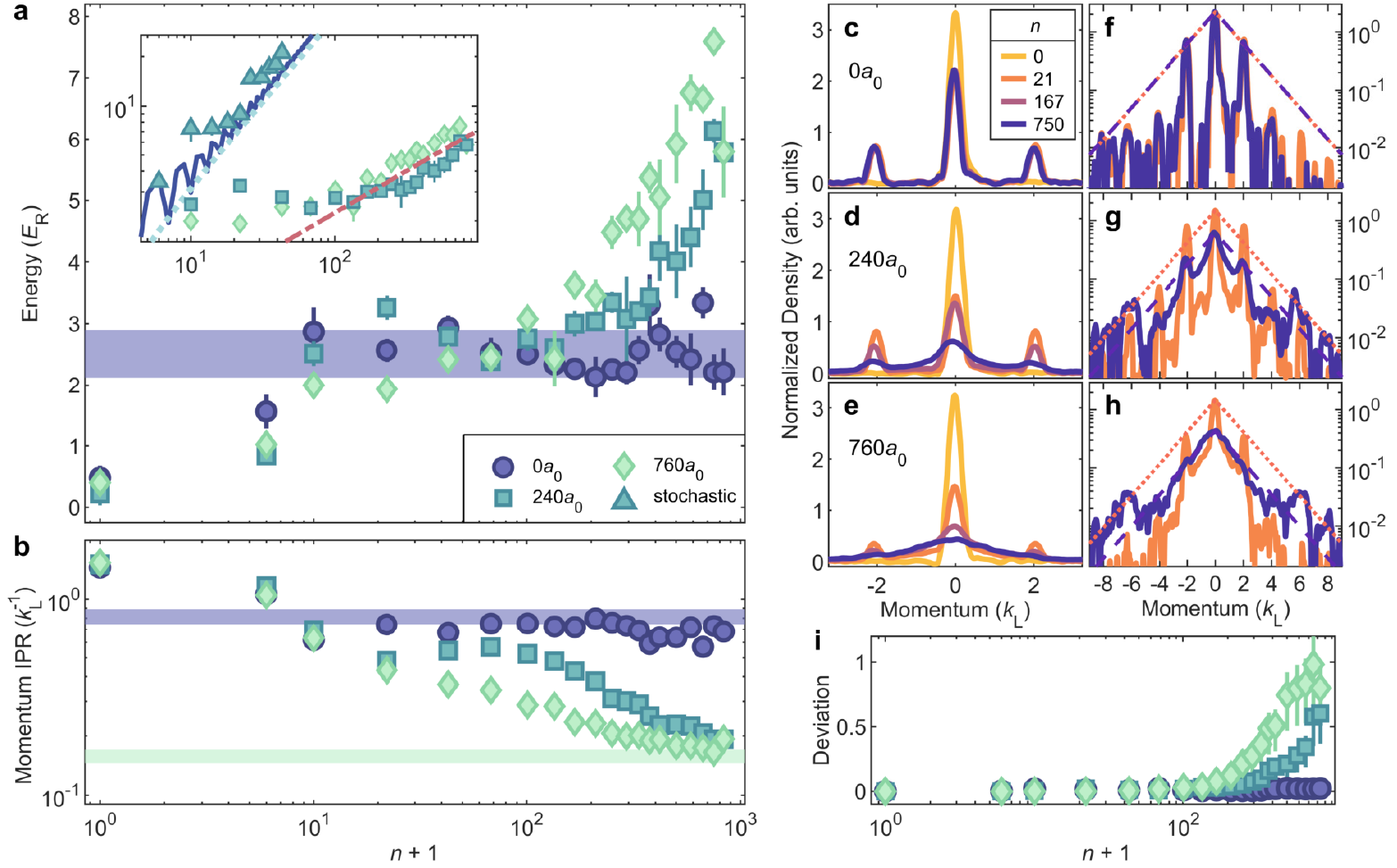}
    \caption{\textbf{Observing the interaction-induced emergence of quantum chaos.} (\textbf{a}) Energy versus kick number for varying $a$. Blue horizontal shaded region indicates the measured single-rotor localization energy of $E_{\mathrm{loc}} = 2.5(4) E_{\mathrm{R}}$. Here $V_0=64E_{\mathrm{R}}$, $T=1.2\,\upmu$s and $\tau=300$ ns ($K \approx 2.3$ and $\kbar \approx 1.5$). The inset contrasts interaction-induced delocalization and anomalous diffusion with classical diffusion caused by random kicking, the latter achieved by adding random offsets to the average kick spacing $T$ drawn uniformly from the interval $\left[-T/4, T/4\right]$. The solid curve is noninteracting quantum theory and the dotted line is a diffusion curve $4 D n/\kbar^2$ with $D \approx 0.19$ extracted from the classical standard map~\cite{Chirikov_CKR_2}. The red dot-dashed line is a subdiffusive $\sqrt{n}$ law serving as a guide to the eye. (\textbf{b}) Momentum-space IPR with transverse dimensions integrated out. The shaded regions are predictions for two exponentially localized distributions with $1/e$ localization length $k_{\mathrm{loc}} = \sqrt{E_{\mathrm{loc}}} \approx 1.6(1) k_{\mathrm{L}}$~\cite{supp}. (\textbf{c-e}) Normalized smoothed momentum space densities at various $n$. (\textbf{f-h}) The same densities on a logarithmic scale. The orange dotted and purple dashed lines are exponentially localized curves $\exp(-k/k_{\mathrm{loc}})$ with amplitudes normalized to match the peak of the measured distributions at the given $n$. (\textbf{i}) Deviation from exponential localization over time based on integrated ratio between measured and exponential distributions with error bars computed from uncertainty in $k_{\mathrm{loc}}$~\cite{supp}.}
    \label{fig:delocalization}
\end{figure*}

% An exciting avenue for exploring the above phenomena is to introduce external periodic driving as well as tunable interactions to a quantum ensemble. The naive expectation is that either will cause heating, but recent theoretical and experimental work has shown the possibility of (dynamically) localized phases and nonergodic behavior when one or both parameters are appropriately controlled.

The quantum kicked rotor (QKR)~\cite{raizen_qkr,quasiperiodicqkr_3dmetaltoinsulator_exp,2qkr_delocalization_exp} is a paradigmatic model of dynamical ergodicity breaking. While strong, repeated kicking drives a classical rotor into chaotic diffusion, the corresponding quantum rotor stops absorbing energy after a finite time, signaling the onset of dynamical localization. Despite the complete absence of disorder, this phenomenon can be understood as a manifestation of Anderson localization in momentum space~\cite{fishman_dynamicallocalization,fishman_dynamicallocalization2}. Although the QKR is a natural context for experimental probes of the interplay between many-body interactions and dynamical localization, the quantum thermodynamics of interacting kicked rotors have not previously been experimentally explored.  %While numerical treatments suggest a breakdown of dynamical localization into anomalous diffusion~\cite{shepelyansky_gpe_qkr_delocalization,manyqkr_delocalization,garreau_meanfield,infcoupledrotor_chaossubdiffusion}, the mean-field approximation on which they are based may be strongly violated by collisional redistribution and quantum depletion~\cite{BECqkr_ring_numerics}, while certain other non-perturbative 1D models display ergodicity-breaking and an MBDL phase~\cite{victor_linearrotors_mbdl,victor_liebliniger_qkr_mbdl}.
Depending on how interactions are introduced into the model, theoretical studies have predicted a variety of novel dynamical phenomena ranging from anomalous diffusion~\cite{shepelyansky_gpe_qkr_delocalization,manyqkr_delocalization,garreau_meanfield,infcoupledrotor_chaossubdiffusion} to classical prethermalization~\cite{delaTorre_classicalprethermal_qkr} to many-body dynamical localization~\cite{victor_liebliniger_qkr_mbdl,victor_linearrotors_mbdl}. 

% The exciting possibilities on both fronts motivate detailed experimental investigation.

Here we report the first experimental study of dynamical localization in the presence of tunable contact interactions, which are nonlocal in momentum space. %Measuring the results of QKR sequences up to one thousand kicks, we observe signatures of a prethermal dynamically-localized regime, followed by interaction-induced anomalous diffusion in momentum space after a variable break time. The role of interactions in destroying reversibility is established using a Loschmidt echo protocol, probing the emergence of the many-body chaotic regime~\cite{loschmidtcooling_exp,loschmidtcooling_theory}. 
These experiments investigate a $^7$Li Bose-Einstein condensate (BEC)  kicked $n$ times at period $T$ by a far-detuned optical lattice of spacing $d=532$ nm and depth $V_0$ for duration $\tau$ (see Fig.~\ref{fig:setup}). We report momentum and energy in units of $k_{\mathrm{L}}=\pi/d$ and $E_{\mathrm{R}}=\hbar^2 k_{\mathrm{L}}^2/2m$ with $m$ the mass of $^7$Li. The single-particle QKR is defined by the 1-cycle Floquet map $U = e^{-i \kbar k^2/2} e^{-iK \cos z/\kbar}$ describing a sharp cosine potential impulse followed by free evolution. Here $k$ and $z$ are momentum and position respectively, $K=\kbar V_0\tau/2\hbar$ is the stochasticity parameter characterizing kicking strength and $\kbar = 8 E_{\mathrm{R}} T/\hbar$ is an effective Planck's constant determined by the kick period.  Absorption imaging after free expansion is used to measure the momentum distribution; see the supplementary text for a full analysis of systematic effects in this procedure, such as dynamics transverse to the lattice direction. Interatomic interactions are varied by tuning the $s$-wave scattering length $a$ (reported in units of the Bohr radius $a_0$) using a magnetic Feshbach resonance. While the kicking primarily couples discrete momentum states along a single dimension, the atoms are entirely unconfined between kicks; scattering between momentum modes thus couples the system to a bath of transverse free-particle states.

The main result of this work is presented in Fig.~\ref{fig:delocalization}a. While a noninteracting sample exhibits dynamical localization, saturating to a finite energy for over 800 kicks, interacting samples clearly demonstrate the destruction of the dynamically localized plateau with increasing scattering length. At intermediate interaction strength ($a=240 a_0$), we observe saturation to the same energy as non-interacting samples for approximately 300 kicks, suggesting the existence of a reasonably long-lived prethermal state. In contrast, the 760$a_0$ trace exceeds this localized energy after around 100 kicks; whether a quasiequilibrium dynamical state is truly established in this stronger-interacting sample is less clear. Fig.~\ref{fig:delocalization}b shows another aspect of the same evolution, plotting the momentum space inverse participation ratio (IPR) versus kick number. The IPR characterizes the number of states over which the system is distributed, thereby also probing how collisional momentum redistribution washes out the originally discrete momentum modes, a process less easily inferred from energy measurements. While the 240$a_0$ data exhibit a clear steady-state behavior for 100 kicks, the 760$a_0$ IPR decreases monotonically for almost the entire experiment. 

A second key result of these measurements is that the observed delocalizing dynamics clearly exhibit anomalous diffusion: it appears that even interacting quantum kicked rotors absorb energy much more slowly than classical rotors. The inset of Fig.~\ref{fig:delocalization}a compares the nature of the observed interaction-induced subdiffusive delocalization with linear energy growth in the classically chaotic model. We experimentally simulate classical dynamics by adding stochastic fluctuations to the kicking period $T$, making use of the known sensitivity of dynamical localization to timing noise~\cite{timingnoise_dl_steckraizen}. These experimental data agree both with single-particle quantum numerics and with the linear energy growth predicted by the classical standard map~\cite{Chirikov_CKR_2}, and stand in clear contrast to the measured interaction-induced anomalous diffusion away from the dynamically localized state. The dot-dashed red line indicates a $\sqrt{n}$ energy growth, and fitting the late-time data to $n^{\alpha}$ yields anomalous diffusion exponents $\alpha$ in the range $\left[0.4,0.6\right]$. For reference, 1D Gross-Pitaevskii simulations on a ring~\cite{garreau_meanfield} predict $\alpha\in\left[0.5,0.8\right]$, though  a direct quantitative comparison to theory is challenging due to the significant condensate depletion and the three-dimensional nature of the experiment. Theoretical studies of the effect of local nonlinearity on real-space Anderson localization instead suggest $\alpha\in\left[ 0.3,0.4 \right]$~\cite{shepelyasnky_gpe_anderson_delocalization,subdiffusion_nonlinear_anderson_theory}, but the long-range nature of contact interactions in momentum space similarly complicates comparison. This clear observation of anomalous diffusion in the interacting quantum kicked rotor raises a variety of fascinating questions for future exploration. What correlations are responsible for the anomalous diffusion dynamics? What feature of the interacting system prevents the interacting QKR from heating classically? What theoretical framework is appropriate for quantitatively predicting wavefunction evolution in this regime? What are the universality properties of the subdiffusive exponent? 

\begin{figure}[!t]
    \centering
    \renewcommand{\figurename}{Fig.}
    \includegraphics[scale = 1]{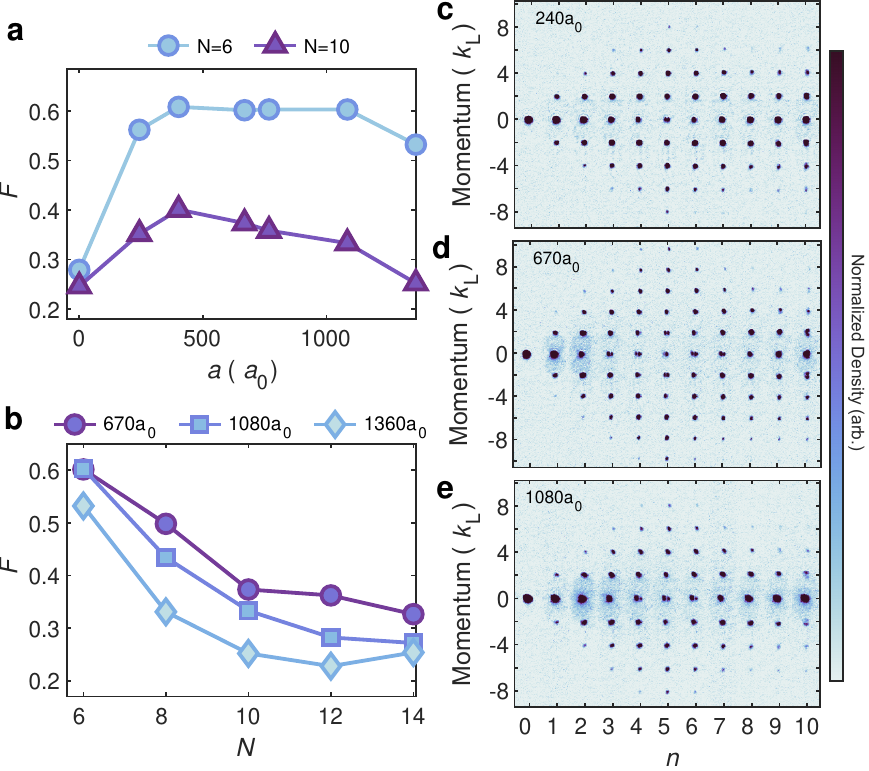}
    \caption{\textbf{Effect of interactions on reversibility in Loschmidt echo experiments.} (\textbf{a}) Measured Loschmidt echo fidelity $F$ for a range of scattering lengths $a=\left[0,1500\right]a_0$ for $N=6$ (blue circles) and $N=10$ (purple triangles), where $N$ indicates the total number of kicks; a first set of $N/2$ kicks propagates the system \textit{forward} in time and a second time-reversal set of $N/2$ kicks propagates it \textit{backwards}.  (\textbf{b}) Measured fidelity $F$ at three different interaction strengths as a function of total number of kicks $N$ in a Loschmidt echo experiment. (\textbf{c-e}) Averaged  absorption images of a BEC after the first $n$ kicks of an $N=10$ Loschmidt echo protocol, for three different $a$.}
    \label{fig:loschmidt}
\end{figure}

% \begin{figure*}
%     \centering
%     \renewcommand{\figurename}{Fig.}
%     \includegraphics[scale=1]{figkickedXXZ_science_withr_v2.pdf}
%     \caption{\textbf{Kicked-rotor dynamics and many-body dynamical localization in kicked spin chains.} (\textbf{A-B}) Calculated single-spin-flip evolution for static and kicked spin-1/2 XXX chains of length $L=13$ for $K=2.4$, $\kbar=1.5$ and $\beta=0$. (\textbf{C}) Spin chain quantum resonance spectroscopy (solid) demonstrating equivalence to QKR (dashed) in the single spin-flip sector with $K=3\kbar$, $L=51$ and averaging over a Gaussian ensemble of $\beta$. Energy is after 10 kicks. Zoomed-in panel compares experimental measurements on the atomic quantum kicked rotor to predictions of spin chain numerics. Dashed vertical lines indicate $q=$ 1/3, 3/8, 2/5, and 1/2. (\textbf{D-E}) The same as A-B but with 2 spin flips. (\textbf{F}) Time average of the staggered magnetization $\langle \sum_j (-1)^j \sigma_j^z \rangle/L$ versus kick number, starting from an initial N\'eel state with $\kbar=1$, $L=12$, $\beta=0.1$ and varying $K$. (\textbf{G}) Staggered magnetization in the infinite-time limit versus $K$, for the same parameters as panel F. (\textbf{H}) Gap-ratio statistic of the $M=0$ sector at $\kbar=1$ for varying $L$ averaged over 100 values of $\beta$. Dashed lines indicate predictions of the Poisson ($\langle r \rangle \approx 0.386$) and circular-orthogonal ensemble ($\langle r \rangle \approx 0.527$)~\cite{floquetthermalization_rigol}.}
%     \label{fig:xxz}
% \end{figure*}

For further insight into the  dynamics of kicked interacting quantum systems we examine the evolution of the momentum distribution, shown in Figs.~\ref{fig:delocalization}c-e. We observe a clear distinction between the noninteracting samples, which settle at a sharply-peaked dynamically-localized momentum distribution, and the interacting samples, which gradually smear out in momentum space due to scattering. Plotting these same densities on a logarithmic scale in Fig.~\ref{fig:delocalization}f-h illuminates the destruction of dynamical Anderson localization by assessing the departure from exponentially-localized Floquet states. The smeared-out lower-energy modes actually appear to maintain the expected localization length, and thus do not trivially indicate a departure from exponential localization. This observation is also reflected in the fact that two predictions based on exponentially localized distributions bound the measured IPR in Fig.~\ref{fig:delocalization}b. Instead, the departure from exponential localization manifests in the emergence of increased relative population in the tails of the distribution. It is interesting to note that recent theory suggests that even many-body dynamically localized phases are expected to exhibit universal power-law decaying tails~\cite{MBDL_powerdecay_rancon}. In Fig.~\ref{fig:delocalization}i we quantitatively characterize the overall deviation from exponential localization~\cite{supp}, revealing a break time near 200 kicks for both interaction strengths. These findings provide a second experimental signature of the destruction of the dynamically localized state by interactions, now both at the level of macroscopic observables and squared wavefunctions.

The onset of energy delocalization due to interactions indicates a transition to the regime of many-body quantum chaos, which can be probed directly by studying time-reversal dynamics~\cite{lyapunovotocqkr_victor,loschmidt_review}. In Fig.~\ref{fig:loschmidt} we probe the onset of chaotic dynamics by measuring the effect of interactions on a Loschmidt echo time-reversal protocol~\cite{loschmidtcooling_exp,loschmidtcooling_theory}. The echo is realized using quantum resonances~\cite{positionspace_quantumresonance} which occur for $\kbar=2\pi q$ with $q$ rational; in particular for $q=2$ ($T \approx 9.95$ $\upmu$s), the free evolution in $U$ largely vanishes and effective time reversal can be achieved by setting $q$ to 1 for a single kick halfway through the sequence~\cite{loschmidtcooling_exp,loschmidtcooling_theory,supp}. This procedure would create exact time reversal for a single zero-quasimomentum state in the absence of interactions.  Due to finite quasimomentum spread and non-reversed interactions, the reversal is imperfect, yielding a Loschmidt fidelity $F = \abs{\bra{\psi} U_2^{\dagger} U_1 \ket{\psi}}^2$ where $U_1$ and $U_2$ are time-evolution operators differing by some perturbation. Perhaps surprisingly, $F$ initially \emph{increases} as the scattering length $a$ is turned up from zero. In this regime $U_1$ and $U_2$ are primarily distinguished by the failure to reverse kinetic energy, and thus the increase can be explained by Thomas-Fermi reduction of the initial state momentum spread. Eventually, for large enough $a$, the interaction becomes the primary perturbation and $F$ begins to decrease with $a$, marking the transition to predominantly interaction-induced irreversibility. The decay of fidelity with total number of kicks in a Loschmidt echo experiment is shown in Fig.~\ref{fig:loschmidt}b. The use of Loschmidt echo techniques as a probe of many-body quantum chaos not only illuminates the origins of the delocalizing dynamics we observe, but opens up the possibility of extending these protocols to probe scrambling in many-body quantum chaotic systems~\cite{otoc2016PRA}.

In conclusion, we have experimentally realized a many-body ensemble of quantum kicked rotors. Following the evolution of interacting samples over hundreds of kicks, we observe signatures of an initial prethermal state, followed by an interaction-induced breakdown of dynamical localization via anomalous diffusion, signaling the onset of many-body quantum chaos. Characterization of the departure from the dynamically localized state indicates subdiffusive energy growth with an exponent near 0.5, easily distinguishable from classical Joule heating in a randomly kicked system, and reveals momentum space distributions which are not exponentially localized. Measuring Loschmidt echo time-reversal dynamics with a quantum resonance enabled us to directly probe the role of interaction-induced irreversibility in driving a transition to many-body quantum chaos. 
% Finally, we have experimentally verified a mapping between the quantum kicked rotor and kicked spin chains in the single-particle limit, and presented numerical evidence for a many-body dynamically localized phase in the latter. 
Together, these results demonstrate and quantitatively illuminate the emergence of interaction-driven quantum chaos in a paradigmatic localized system. %and pave the way for the exploration and application of many-body dynamical localization and disorder-free dynamic stabilization in a broad range of physical contexts.

We acknowledge helpful conversations with Adam Ran{\c{c}}on, Norman Yao, and Thomas Schuster. During the performance of the research described here we became aware of related efforts underway in another experimental group~\cite{interactingqkr_deep} which have reached similar conclusions using a complementary approach. 
\textbf{Funding:} D.M.W. acknowledges support from the Air Force Office of Scientific Research (AFOSR FA9550-20-1-0240), the Army Research Office (PECASE W911NF1410154), and the National Science Foundation (CAREER 1555313 and QLCI OMA-2016245). D.M.W., R.S., and E.N-M. acknowledge support from the UCSB NSF Quantum Foundry through the Q-AMASE-i program (Grant No. DMR-1906325). V.G. was supported by US-ARO Contract No.W911NF1310172, NSF DMR-2037158, and the Simons Foundation. \textbf{Authors' contributions:} A.C., R.S., H.M., E.Q.S., J.L.T., E.N-M., T.S. and H.E.K. contributed to operating the experiment and performing the measurements. A.C., R.S. and H.M. analyzed the data. A.C. conceptualized and performed the theoretical simulations of rotor and spin models. V.G. and D.M.W. developed the idea for the experiment. D.M.W. supervised the work. A.C., R.S., H.M., V.G. and D.M.W. wrote the manuscript. All authors contributed to discussion and interpretation of the results. \textbf{Competing interests:} The authors declare no competing interests. \textbf{Data and materials availability:} All data needed to evaluate the conclusions in this study are presented in the paper and in the supplementary materials.

\let\oldaddcontentsline\addcontentsline
\renewcommand{\addcontentsline}[3]{}
\bibliography{qkrbib}

\begin{thebibliography}{54}%
\makeatletter
\providecommand \@ifxundefined [1]{%
 \@ifx{#1\undefined}
}%
\providecommand \@ifnum [1]{%
 \ifnum #1\expandafter \@firstoftwo
 \else \expandafter \@secondoftwo
 \fi
}%
\providecommand \@ifx [1]{%
 \ifx #1\expandafter \@firstoftwo
 \else \expandafter \@secondoftwo
 \fi
}%
\providecommand \natexlab [1]{#1}%
\providecommand \enquote  [1]{``#1''}%
\providecommand \bibnamefont  [1]{#1}%
\providecommand \bibfnamefont [1]{#1}%
\providecommand \citenamefont [1]{#1}%
\providecommand \href@noop [0]{\@secondoftwo}%
\providecommand \href [0]{\begingroup \@sanitize@url \@href}%
\providecommand \@href[1]{\@@startlink{#1}\@@href}%
\providecommand \@@href[1]{\endgroup#1\@@endlink}%
\providecommand \@sanitize@url [0]{\catcode `\\12\catcode `\$12\catcode
  `\&12\catcode `\#12\catcode `\^12\catcode `\_12\catcode `\%12\relax}%
\providecommand \@@startlink[1]{}%
\providecommand \@@endlink[0]{}%
\providecommand \url  [0]{\begingroup\@sanitize@url \@url }%
\providecommand \@url [1]{\endgroup\@href {#1}{\urlprefix }}%
\providecommand \urlprefix  [0]{URL }%
\providecommand \Eprint [0]{\href }%
\providecommand \doibase [0]{https://doi.org/}%
\providecommand \selectlanguage [0]{\@gobble}%
\providecommand \bibinfo  [0]{\@secondoftwo}%
\providecommand \bibfield  [0]{\@secondoftwo}%
\providecommand \translation [1]{[#1]}%
\providecommand \BibitemOpen [0]{}%
\providecommand \bibitemStop [0]{}%
\providecommand \bibitemNoStop [0]{.\EOS\space}%
\providecommand \EOS [0]{\spacefactor3000\relax}%
\providecommand \BibitemShut  [1]{\csname bibitem#1\endcsname}%
\let\auto@bib@innerbib\@empty
%</preamble>
\bibitem [{\citenamefont {Nandkishore}\ and\ \citenamefont
  {Huse}(2015)}]{mblreview_huse}%
  \BibitemOpen
  \bibfield  {author} {\bibinfo {author} {\bibfnamefont {R.}~\bibnamefont
  {Nandkishore}}\ and\ \bibinfo {author} {\bibfnamefont {D.~A.}\ \bibnamefont
  {Huse}},\ }\href {https://doi.org/10.1146/annurev-conmatphys-031214-014726}
  {\bibfield  {journal} {\bibinfo  {journal} {Annu. Rev. Condens. Matter
  Phys.}\ }\textbf {\bibinfo {volume} {6}},\ \bibinfo {pages} {15} (\bibinfo
  {year} {2015})}\BibitemShut {NoStop}%
\bibitem [{\citenamefont {Abanin}\ \emph {et~al.}(2019)\citenamefont {Abanin},
  \citenamefont {Altman}, \citenamefont {Bloch},\ and\ \citenamefont
  {Serbyn}}]{mblreview_abanin}%
  \BibitemOpen
  \bibfield  {author} {\bibinfo {author} {\bibfnamefont {D.~A.}\ \bibnamefont
  {Abanin}}, \bibinfo {author} {\bibfnamefont {E.}~\bibnamefont {Altman}},
  \bibinfo {author} {\bibfnamefont {I.}~\bibnamefont {Bloch}},\ and\ \bibinfo
  {author} {\bibfnamefont {M.}~\bibnamefont {Serbyn}},\ }\href
  {https://doi.org/10.1103/RevModPhys.91.021001} {\bibfield  {journal}
  {\bibinfo  {journal} {Rev. Mod. Phys.}\ }\textbf {\bibinfo {volume} {91}},\
  \bibinfo {pages} {021001} (\bibinfo {year} {2019})}\BibitemShut {NoStop}%
\bibitem [{\citenamefont {Bernien}\ \emph {et~al.}(2017)\citenamefont
  {Bernien}, \citenamefont {Schwartz}, \citenamefont {Keesling}, \citenamefont
  {Levine}, \citenamefont {Omran}, \citenamefont {Pichler}, \citenamefont
  {Choi}, \citenamefont {Zibrov}, \citenamefont {Endres}, \citenamefont
  {Greiner}, \citenamefont {Vuletić},\ and\ \citenamefont
  {Lukin}}]{bernien_scars_2017}%
  \BibitemOpen
  \bibfield  {author} {\bibinfo {author} {\bibfnamefont {H.}~\bibnamefont
  {Bernien}}, \bibinfo {author} {\bibfnamefont {S.}~\bibnamefont {Schwartz}},
  \bibinfo {author} {\bibfnamefont {A.}~\bibnamefont {Keesling}}, \bibinfo
  {author} {\bibfnamefont {H.}~\bibnamefont {Levine}}, \bibinfo {author}
  {\bibfnamefont {A.}~\bibnamefont {Omran}}, \bibinfo {author} {\bibfnamefont
  {H.}~\bibnamefont {Pichler}}, \bibinfo {author} {\bibfnamefont
  {S.}~\bibnamefont {Choi}}, \bibinfo {author} {\bibfnamefont {A.~S.}\
  \bibnamefont {Zibrov}}, \bibinfo {author} {\bibfnamefont {M.}~\bibnamefont
  {Endres}}, \bibinfo {author} {\bibfnamefont {M.}~\bibnamefont {Greiner}},
  \bibinfo {author} {\bibfnamefont {V.}~\bibnamefont {Vuletić}},\ and\
  \bibinfo {author} {\bibfnamefont {M.~D.}\ \bibnamefont {Lukin}},\ }\href
  {https://doi.org/10.1038/nature24622} {\bibfield  {journal} {\bibinfo
  {journal} {Nature}\ }\textbf {\bibinfo {volume} {551}},\ \bibinfo {pages}
  {579} (\bibinfo {year} {2017})}\BibitemShut {NoStop}%
\bibitem [{\citenamefont {Serbyn}\ \emph {et~al.}(2021)\citenamefont {Serbyn},
  \citenamefont {Abanin},\ and\ \citenamefont {Papić}}]{scarsreview_2021}%
  \BibitemOpen
  \bibfield  {author} {\bibinfo {author} {\bibfnamefont {M.}~\bibnamefont
  {Serbyn}}, \bibinfo {author} {\bibfnamefont {D.~A.}\ \bibnamefont {Abanin}},\
  and\ \bibinfo {author} {\bibfnamefont {Z.}~\bibnamefont {Papić}},\ }\href
  {https://doi.org/10.1038/s41567-021-01230-2} {\bibfield  {journal} {\bibinfo
  {journal} {Nature Physics}\ }\textbf {\bibinfo {volume} {17}},\ \bibinfo
  {pages} {675} (\bibinfo {year} {2021})}\BibitemShut {NoStop}%
\bibitem [{\citenamefont {Singh}\ \emph {et~al.}(2019)\citenamefont {Singh},
  \citenamefont {Fujiwara}, \citenamefont {Geiger}, \citenamefont {Simmons},
  \citenamefont {Lipatov}, \citenamefont {Cao}, \citenamefont {Dotti},
  \citenamefont {Rajagopal}, \citenamefont {Senaratne}, \citenamefont
  {Shimasaki}, \citenamefont {Heyl}, \citenamefont {Eckardt},\ and\
  \citenamefont {Weld}}]{prethermal_weld}%
  \BibitemOpen
  \bibfield  {author} {\bibinfo {author} {\bibfnamefont {K.}~\bibnamefont
  {Singh}}, \bibinfo {author} {\bibfnamefont {C.~J.}\ \bibnamefont {Fujiwara}},
  \bibinfo {author} {\bibfnamefont {Z.~A.}\ \bibnamefont {Geiger}}, \bibinfo
  {author} {\bibfnamefont {E.~Q.}\ \bibnamefont {Simmons}}, \bibinfo {author}
  {\bibfnamefont {M.}~\bibnamefont {Lipatov}}, \bibinfo {author} {\bibfnamefont
  {A.}~\bibnamefont {Cao}}, \bibinfo {author} {\bibfnamefont {P.}~\bibnamefont
  {Dotti}}, \bibinfo {author} {\bibfnamefont {S.~V.}\ \bibnamefont
  {Rajagopal}}, \bibinfo {author} {\bibfnamefont {R.}~\bibnamefont
  {Senaratne}}, \bibinfo {author} {\bibfnamefont {T.}~\bibnamefont
  {Shimasaki}}, \bibinfo {author} {\bibfnamefont {M.}~\bibnamefont {Heyl}},
  \bibinfo {author} {\bibfnamefont {A.}~\bibnamefont {Eckardt}},\ and\ \bibinfo
  {author} {\bibfnamefont {D.~M.}\ \bibnamefont {Weld}},\ }\href
  {https://doi.org/10.1103/PhysRevX.9.041021} {\bibfield  {journal} {\bibinfo
  {journal} {Phys. Rev. X}\ }\textbf {\bibinfo {volume} {9}},\ \bibinfo {pages}
  {041021} (\bibinfo {year} {2019})}\BibitemShut {NoStop}%
\bibitem [{\citenamefont {Rubio-Abadal}\ \emph {et~al.}(2020)\citenamefont
  {Rubio-Abadal}, \citenamefont {Ippoliti}, \citenamefont {Hollerith},
  \citenamefont {Wei}, \citenamefont {Rui}, \citenamefont {Sondhi},
  \citenamefont {Khemani}, \citenamefont {Gross},\ and\ \citenamefont
  {Bloch}}]{prethermal_bloch}%
  \BibitemOpen
  \bibfield  {author} {\bibinfo {author} {\bibfnamefont {A.}~\bibnamefont
  {Rubio-Abadal}}, \bibinfo {author} {\bibfnamefont {M.}~\bibnamefont
  {Ippoliti}}, \bibinfo {author} {\bibfnamefont {S.}~\bibnamefont {Hollerith}},
  \bibinfo {author} {\bibfnamefont {D.}~\bibnamefont {Wei}}, \bibinfo {author}
  {\bibfnamefont {J.}~\bibnamefont {Rui}}, \bibinfo {author} {\bibfnamefont
  {S.~L.}\ \bibnamefont {Sondhi}}, \bibinfo {author} {\bibfnamefont
  {V.}~\bibnamefont {Khemani}}, \bibinfo {author} {\bibfnamefont
  {C.}~\bibnamefont {Gross}},\ and\ \bibinfo {author} {\bibfnamefont
  {I.}~\bibnamefont {Bloch}},\ }\href
  {https://doi.org/10.1103/PhysRevX.10.021044} {\bibfield  {journal} {\bibinfo
  {journal} {Phys. Rev. X}\ }\textbf {\bibinfo {volume} {10}},\ \bibinfo
  {pages} {021044} (\bibinfo {year} {2020})}\BibitemShut {NoStop}%
\bibitem [{\citenamefont {Peng}\ \emph {et~al.}(2021)\citenamefont {Peng},
  \citenamefont {Yin}, \citenamefont {Huang}, \citenamefont {Ramanathan},\ and\
  \citenamefont {Cappellaro}}]{prethermal_cappellaro}%
  \BibitemOpen
  \bibfield  {author} {\bibinfo {author} {\bibfnamefont {P.}~\bibnamefont
  {Peng}}, \bibinfo {author} {\bibfnamefont {C.}~\bibnamefont {Yin}}, \bibinfo
  {author} {\bibfnamefont {X.}~\bibnamefont {Huang}}, \bibinfo {author}
  {\bibfnamefont {C.}~\bibnamefont {Ramanathan}},\ and\ \bibinfo {author}
  {\bibfnamefont {P.}~\bibnamefont {Cappellaro}},\ }\href@noop {} {\bibfield
  {journal} {\bibinfo  {journal} {Nat. Phys.}\ } (\bibinfo {year}
  {2021})}\BibitemShut {NoStop}%
\bibitem [{\citenamefont {Langen}\ \emph {et~al.}(2015)\citenamefont {Langen},
  \citenamefont {Erne}, \citenamefont {Geiger}, \citenamefont {Rauer},
  \citenamefont {Schweigler}, \citenamefont {Kuhnert}, \citenamefont
  {Rohringer}, \citenamefont {Mazets}, \citenamefont {Gasenzer},\ and\
  \citenamefont {Schmiedmayer}}]{prethermal_schmiedmayer}%
  \BibitemOpen
  \bibfield  {author} {\bibinfo {author} {\bibfnamefont {T.}~\bibnamefont
  {Langen}}, \bibinfo {author} {\bibfnamefont {S.}~\bibnamefont {Erne}},
  \bibinfo {author} {\bibfnamefont {R.}~\bibnamefont {Geiger}}, \bibinfo
  {author} {\bibfnamefont {B.}~\bibnamefont {Rauer}}, \bibinfo {author}
  {\bibfnamefont {T.}~\bibnamefont {Schweigler}}, \bibinfo {author}
  {\bibfnamefont {M.}~\bibnamefont {Kuhnert}}, \bibinfo {author} {\bibfnamefont
  {W.}~\bibnamefont {Rohringer}}, \bibinfo {author} {\bibfnamefont {I.~E.}\
  \bibnamefont {Mazets}}, \bibinfo {author} {\bibfnamefont {T.}~\bibnamefont
  {Gasenzer}},\ and\ \bibinfo {author} {\bibfnamefont {J.}~\bibnamefont
  {Schmiedmayer}},\ }\href {https://doi.org/10.1126/science.1257026} {\bibfield
   {journal} {\bibinfo  {journal} {Science}\ }\textbf {\bibinfo {volume}
  {348}},\ \bibinfo {pages} {207} (\bibinfo {year} {2015})}\BibitemShut
  {NoStop}%
\bibitem [{\citenamefont {Ueda}(2020)}]{ueda_prethermalreview_2020}%
  \BibitemOpen
  \bibfield  {author} {\bibinfo {author} {\bibfnamefont {M.}~\bibnamefont
  {Ueda}},\ }\href {https://doi.org/10.1038/s42254-020-0237-x} {\bibfield
  {journal} {\bibinfo  {journal} {Nature Reviews Physics}\ }\textbf {\bibinfo
  {volume} {2}},\ \bibinfo {pages} {669} (\bibinfo {year} {2020})}\BibitemShut
  {NoStop}%
\bibitem [{\citenamefont {Agarwal}\ \emph {et~al.}(2015)\citenamefont
  {Agarwal}, \citenamefont {Gopalakrishnan}, \citenamefont {Knap},
  \citenamefont {M\"uller},\ and\ \citenamefont
  {Demler}}]{mblsubdiffusion_demler}%
  \BibitemOpen
  \bibfield  {author} {\bibinfo {author} {\bibfnamefont {K.}~\bibnamefont
  {Agarwal}}, \bibinfo {author} {\bibfnamefont {S.}~\bibnamefont
  {Gopalakrishnan}}, \bibinfo {author} {\bibfnamefont {M.}~\bibnamefont
  {Knap}}, \bibinfo {author} {\bibfnamefont {M.}~\bibnamefont {M\"uller}},\
  and\ \bibinfo {author} {\bibfnamefont {E.}~\bibnamefont {Demler}},\ }\href
  {https://doi.org/10.1103/PhysRevLett.114.160401} {\bibfield  {journal}
  {\bibinfo  {journal} {Phys. Rev. Lett.}\ }\textbf {\bibinfo {volume} {114}},\
  \bibinfo {pages} {160401} (\bibinfo {year} {2015})}\BibitemShut {NoStop}%
\bibitem [{\citenamefont {Weidinger}\ and\ \citenamefont
  {Knap}(2017)}]{prethermal_subjoule_ON_knap}%
  \BibitemOpen
  \bibfield  {author} {\bibinfo {author} {\bibfnamefont {S.~A.}\ \bibnamefont
  {Weidinger}}\ and\ \bibinfo {author} {\bibfnamefont {M.}~\bibnamefont
  {Knap}},\ }\href@noop {} {\bibfield  {journal} {\bibinfo  {journal} {Sci.
  Rep.}\ }\textbf {\bibinfo {volume} {7}},\ \bibinfo {pages} {1} (\bibinfo
  {year} {2017})}\BibitemShut {NoStop}%
\bibitem [{\citenamefont {Menu}\ and\ \citenamefont
  {Roscilde}(2020)}]{anomalousspin}%
  \BibitemOpen
  \bibfield  {author} {\bibinfo {author} {\bibfnamefont {R.}~\bibnamefont
  {Menu}}\ and\ \bibinfo {author} {\bibfnamefont {T.}~\bibnamefont
  {Roscilde}},\ }\href {https://doi.org/10.1103/PhysRevLett.124.130604}
  {\bibfield  {journal} {\bibinfo  {journal} {Phys. Rev. Lett.}\ }\textbf
  {\bibinfo {volume} {124}},\ \bibinfo {pages} {130604} (\bibinfo {year}
  {2020})}\BibitemShut {NoStop}%
\bibitem [{sup()}]{supp}%
  \BibitemOpen
  \href@noop {} {}\bibinfo {note} {Materials and methods are available as
  supplementary materials.}\BibitemShut {Stop}%
\bibitem [{\citenamefont {Chirikov}(1979)}]{Chirikov_CKR_2}%
  \BibitemOpen
  \bibfield  {author} {\bibinfo {author} {\bibfnamefont {B.~V.}\ \bibnamefont
  {Chirikov}},\ }\href
  {https://doi.org/https://doi.org/10.1016/0370-1573(79)90023-1} {\bibfield
  {journal} {\bibinfo  {journal} {Phys. Rep.}\ }\textbf {\bibinfo {volume}
  {52}},\ \bibinfo {pages} {263 } (\bibinfo {year} {1979})}\BibitemShut
  {NoStop}%
\bibitem [{\citenamefont {Moore}\ \emph {et~al.}(1995)\citenamefont {Moore},
  \citenamefont {Robinson}, \citenamefont {Bharucha}, \citenamefont
  {Sundaram},\ and\ \citenamefont {Raizen}}]{raizen_qkr}%
  \BibitemOpen
  \bibfield  {author} {\bibinfo {author} {\bibfnamefont {F.~L.}\ \bibnamefont
  {Moore}}, \bibinfo {author} {\bibfnamefont {J.~C.}\ \bibnamefont {Robinson}},
  \bibinfo {author} {\bibfnamefont {C.~F.}\ \bibnamefont {Bharucha}}, \bibinfo
  {author} {\bibfnamefont {B.}~\bibnamefont {Sundaram}},\ and\ \bibinfo
  {author} {\bibfnamefont {M.~G.}\ \bibnamefont {Raizen}},\ }\href
  {https://doi.org/10.1103/PhysRevLett.75.4598} {\bibfield  {journal} {\bibinfo
   {journal} {Phys. Rev. Lett.}\ }\textbf {\bibinfo {volume} {75}},\ \bibinfo
  {pages} {4598} (\bibinfo {year} {1995})}\BibitemShut {NoStop}%
\bibitem [{\citenamefont {Chab\'e}\ \emph {et~al.}(2008)\citenamefont
  {Chab\'e}, \citenamefont {Lemari\'e}, \citenamefont {Gr\'emaud},
  \citenamefont {Delande}, \citenamefont {Szriftgiser},\ and\ \citenamefont
  {Garreau}}]{quasiperiodicqkr_3dmetaltoinsulator_exp}%
  \BibitemOpen
  \bibfield  {author} {\bibinfo {author} {\bibfnamefont {J.}~\bibnamefont
  {Chab\'e}}, \bibinfo {author} {\bibfnamefont {G.}~\bibnamefont {Lemari\'e}},
  \bibinfo {author} {\bibfnamefont {B.}~\bibnamefont {Gr\'emaud}}, \bibinfo
  {author} {\bibfnamefont {D.}~\bibnamefont {Delande}}, \bibinfo {author}
  {\bibfnamefont {P.}~\bibnamefont {Szriftgiser}},\ and\ \bibinfo {author}
  {\bibfnamefont {J.~C.}\ \bibnamefont {Garreau}},\ }\href
  {https://doi.org/10.1103/PhysRevLett.101.255702} {\bibfield  {journal}
  {\bibinfo  {journal} {Phys. Rev. Lett.}\ }\textbf {\bibinfo {volume} {101}},\
  \bibinfo {pages} {255702} (\bibinfo {year} {2008})}\BibitemShut {NoStop}%
\bibitem [{\citenamefont {Gadway}\ \emph {et~al.}(2013)\citenamefont {Gadway},
  \citenamefont {Reeves}, \citenamefont {Krinner},\ and\ \citenamefont
  {Schneble}}]{2qkr_delocalization_exp}%
  \BibitemOpen
  \bibfield  {author} {\bibinfo {author} {\bibfnamefont {B.}~\bibnamefont
  {Gadway}}, \bibinfo {author} {\bibfnamefont {J.}~\bibnamefont {Reeves}},
  \bibinfo {author} {\bibfnamefont {L.}~\bibnamefont {Krinner}},\ and\ \bibinfo
  {author} {\bibfnamefont {D.}~\bibnamefont {Schneble}},\ }\href
  {https://doi.org/10.1103/PhysRevLett.110.190401} {\bibfield  {journal}
  {\bibinfo  {journal} {Phys. Rev. Lett.}\ }\textbf {\bibinfo {volume} {110}},\
  \bibinfo {pages} {190401} (\bibinfo {year} {2013})}\BibitemShut {NoStop}%
\bibitem [{\citenamefont {Fishman}\ \emph {et~al.}(1982)\citenamefont
  {Fishman}, \citenamefont {Grempel},\ and\ \citenamefont
  {Prange}}]{fishman_dynamicallocalization}%
  \BibitemOpen
  \bibfield  {author} {\bibinfo {author} {\bibfnamefont {S.}~\bibnamefont
  {Fishman}}, \bibinfo {author} {\bibfnamefont {D.~R.}\ \bibnamefont
  {Grempel}},\ and\ \bibinfo {author} {\bibfnamefont {R.~E.}\ \bibnamefont
  {Prange}},\ }\href {https://doi.org/10.1103/PhysRevLett.49.509} {\bibfield
  {journal} {\bibinfo  {journal} {Phys. Rev. Lett.}\ }\textbf {\bibinfo
  {volume} {49}},\ \bibinfo {pages} {509} (\bibinfo {year} {1982})}\BibitemShut
  {NoStop}%
\bibitem [{\citenamefont {Grempel}\ \emph {et~al.}(1984)\citenamefont
  {Grempel}, \citenamefont {Prange},\ and\ \citenamefont
  {Fishman}}]{fishman_dynamicallocalization2}%
  \BibitemOpen
  \bibfield  {author} {\bibinfo {author} {\bibfnamefont {D.~R.}\ \bibnamefont
  {Grempel}}, \bibinfo {author} {\bibfnamefont {R.~E.}\ \bibnamefont
  {Prange}},\ and\ \bibinfo {author} {\bibfnamefont {S.}~\bibnamefont
  {Fishman}},\ }\href {https://doi.org/10.1103/PhysRevA.29.1639} {\bibfield
  {journal} {\bibinfo  {journal} {Phys. Rev. A}\ }\textbf {\bibinfo {volume}
  {29}},\ \bibinfo {pages} {1639} (\bibinfo {year} {1984})}\BibitemShut
  {NoStop}%
\bibitem [{\citenamefont
  {Shepelyansky}(1993)}]{shepelyansky_gpe_qkr_delocalization}%
  \BibitemOpen
  \bibfield  {author} {\bibinfo {author} {\bibfnamefont {D.~L.}\ \bibnamefont
  {Shepelyansky}},\ }\href {https://doi.org/10.1103/PhysRevLett.70.1787}
  {\bibfield  {journal} {\bibinfo  {journal} {Phys. Rev. Lett.}\ }\textbf
  {\bibinfo {volume} {70}},\ \bibinfo {pages} {1787} (\bibinfo {year}
  {1993})}\BibitemShut {NoStop}%
\bibitem [{\citenamefont {Gligori{\'{c}}}\ \emph {et~al.}(2011)\citenamefont
  {Gligori{\'{c}}}, \citenamefont {Bodyfelt},\ and\ \citenamefont
  {Flach}}]{manyqkr_delocalization}%
  \BibitemOpen
  \bibfield  {author} {\bibinfo {author} {\bibfnamefont {G.}~\bibnamefont
  {Gligori{\'{c}}}}, \bibinfo {author} {\bibfnamefont {J.~D.}\ \bibnamefont
  {Bodyfelt}},\ and\ \bibinfo {author} {\bibfnamefont {S.}~\bibnamefont
  {Flach}},\ }\href {https://doi.org/10.1209/0295-5075/96/30004} {\bibfield
  {journal} {\bibinfo  {journal} {Europhys. Lett.}\ }\textbf {\bibinfo {volume}
  {96}},\ \bibinfo {pages} {30004} (\bibinfo {year} {2011})}\BibitemShut
  {NoStop}%
\bibitem [{\citenamefont {Lellouch}\ \emph {et~al.}(2020)\citenamefont
  {Lellouch}, \citenamefont {Ran\ifmmode~\mbox{\c{c}}\else \c{c}\fi{}on},
  \citenamefont {De~Bi\`evre}, \citenamefont {Delande},\ and\ \citenamefont
  {Garreau}}]{garreau_meanfield}%
  \BibitemOpen
  \bibfield  {author} {\bibinfo {author} {\bibfnamefont {S.}~\bibnamefont
  {Lellouch}}, \bibinfo {author} {\bibfnamefont {A.}~\bibnamefont
  {Ran\ifmmode~\mbox{\c{c}}\else \c{c}\fi{}on}}, \bibinfo {author}
  {\bibfnamefont {S.}~\bibnamefont {De~Bi\`evre}}, \bibinfo {author}
  {\bibfnamefont {D.}~\bibnamefont {Delande}},\ and\ \bibinfo {author}
  {\bibfnamefont {J.~C.}\ \bibnamefont {Garreau}},\ }\href
  {https://doi.org/10.1103/PhysRevA.101.043624} {\bibfield  {journal} {\bibinfo
   {journal} {Phys. Rev. A}\ }\textbf {\bibinfo {volume} {101}},\ \bibinfo
  {pages} {043624} (\bibinfo {year} {2020})}\BibitemShut {NoStop}%
\bibitem [{\citenamefont {Russomanno}\ \emph {et~al.}(2021)\citenamefont
  {Russomanno}, \citenamefont {Fava},\ and\ \citenamefont
  {Fazio}}]{infcoupledrotor_chaossubdiffusion}%
  \BibitemOpen
  \bibfield  {author} {\bibinfo {author} {\bibfnamefont {A.}~\bibnamefont
  {Russomanno}}, \bibinfo {author} {\bibfnamefont {M.}~\bibnamefont {Fava}},\
  and\ \bibinfo {author} {\bibfnamefont {R.}~\bibnamefont {Fazio}},\ }\href
  {https://doi.org/10.1103/PhysRevB.103.224301} {\bibfield  {journal} {\bibinfo
   {journal} {Phys. Rev. B}\ }\textbf {\bibinfo {volume} {103}},\ \bibinfo
  {pages} {224301} (\bibinfo {year} {2021})}\BibitemShut {NoStop}%
\bibitem [{\citenamefont {Rajak}\ \emph {et~al.}(2019)\citenamefont {Rajak},
  \citenamefont {Dana},\ and\ \citenamefont
  {Dalla~Torre}}]{delaTorre_classicalprethermal_qkr}%
  \BibitemOpen
  \bibfield  {author} {\bibinfo {author} {\bibfnamefont {A.}~\bibnamefont
  {Rajak}}, \bibinfo {author} {\bibfnamefont {I.}~\bibnamefont {Dana}},\ and\
  \bibinfo {author} {\bibfnamefont {E.~G.}\ \bibnamefont {Dalla~Torre}},\
  }\href {https://doi.org/10.1103/PhysRevB.100.100302} {\bibfield  {journal}
  {\bibinfo  {journal} {Phys. Rev. B}\ }\textbf {\bibinfo {volume} {100}},\
  \bibinfo {pages} {100302} (\bibinfo {year} {2019})}\BibitemShut {NoStop}%
\bibitem [{\citenamefont {Rylands}\ \emph {et~al.}(2020)\citenamefont
  {Rylands}, \citenamefont {Rozenbaum}, \citenamefont {Galitski},\ and\
  \citenamefont {Konik}}]{victor_liebliniger_qkr_mbdl}%
  \BibitemOpen
  \bibfield  {author} {\bibinfo {author} {\bibfnamefont {C.}~\bibnamefont
  {Rylands}}, \bibinfo {author} {\bibfnamefont {E.~B.}\ \bibnamefont
  {Rozenbaum}}, \bibinfo {author} {\bibfnamefont {V.}~\bibnamefont
  {Galitski}},\ and\ \bibinfo {author} {\bibfnamefont {R.}~\bibnamefont
  {Konik}},\ }\href {https://doi.org/10.1103/PhysRevLett.124.155302} {\bibfield
   {journal} {\bibinfo  {journal} {Phys. Rev. Lett.}\ }\textbf {\bibinfo
  {volume} {124}},\ \bibinfo {pages} {155302} (\bibinfo {year}
  {2020})}\BibitemShut {NoStop}%
\bibitem [{\citenamefont {Keser}\ \emph {et~al.}(2016)\citenamefont {Keser},
  \citenamefont {Ganeshan}, \citenamefont {Refael},\ and\ \citenamefont
  {Galitski}}]{victor_linearrotors_mbdl}%
  \BibitemOpen
  \bibfield  {author} {\bibinfo {author} {\bibfnamefont {A.~C.}\ \bibnamefont
  {Keser}}, \bibinfo {author} {\bibfnamefont {S.}~\bibnamefont {Ganeshan}},
  \bibinfo {author} {\bibfnamefont {G.}~\bibnamefont {Refael}},\ and\ \bibinfo
  {author} {\bibfnamefont {V.}~\bibnamefont {Galitski}},\ }\href
  {https://doi.org/10.1103/PhysRevB.94.085120} {\bibfield  {journal} {\bibinfo
  {journal} {Phys. Rev. B}\ }\textbf {\bibinfo {volume} {94}},\ \bibinfo
  {pages} {085120} (\bibinfo {year} {2016})}\BibitemShut {NoStop}%
\bibitem [{\citenamefont {Oskay}\ \emph {et~al.}(2003)\citenamefont {Oskay},
  \citenamefont {Steck},\ and\ \citenamefont
  {Raizen}}]{timingnoise_dl_steckraizen}%
  \BibitemOpen
  \bibfield  {author} {\bibinfo {author} {\bibfnamefont {W.~H.}\ \bibnamefont
  {Oskay}}, \bibinfo {author} {\bibfnamefont {D.~A.}\ \bibnamefont {Steck}},\
  and\ \bibinfo {author} {\bibfnamefont {M.~G.}\ \bibnamefont {Raizen}},\
  }\href {https://doi.org/https://doi.org/10.1016/S0960-0779(02)00302-8}
  {\bibfield  {journal} {\bibinfo  {journal} {Chaos, Solitons and Fractals}\
  }\textbf {\bibinfo {volume} {16}},\ \bibinfo {pages} {409} (\bibinfo {year}
  {2003})}\BibitemShut {NoStop}%
\bibitem [{\citenamefont {Pikovsky}\ and\ \citenamefont
  {Shepelyansky}(2008)}]{shepelyasnky_gpe_anderson_delocalization}%
  \BibitemOpen
  \bibfield  {author} {\bibinfo {author} {\bibfnamefont {A.~S.}\ \bibnamefont
  {Pikovsky}}\ and\ \bibinfo {author} {\bibfnamefont {D.~L.}\ \bibnamefont
  {Shepelyansky}},\ }\href {https://doi.org/10.1103/PhysRevLett.100.094101}
  {\bibfield  {journal} {\bibinfo  {journal} {Phys. Rev. Lett.}\ }\textbf
  {\bibinfo {volume} {100}},\ \bibinfo {pages} {094101} (\bibinfo {year}
  {2008})}\BibitemShut {NoStop}%
\bibitem [{\citenamefont {Flach}\ \emph {et~al.}(2009)\citenamefont {Flach},
  \citenamefont {Krimer},\ and\ \citenamefont
  {Skokos}}]{subdiffusion_nonlinear_anderson_theory}%
  \BibitemOpen
  \bibfield  {author} {\bibinfo {author} {\bibfnamefont {S.}~\bibnamefont
  {Flach}}, \bibinfo {author} {\bibfnamefont {D.~O.}\ \bibnamefont {Krimer}},\
  and\ \bibinfo {author} {\bibfnamefont {C.}~\bibnamefont {Skokos}},\ }\href
  {https://doi.org/10.1103/PhysRevLett.102.024101} {\bibfield  {journal}
  {\bibinfo  {journal} {Phys. Rev. Lett.}\ }\textbf {\bibinfo {volume} {102}},\
  \bibinfo {pages} {024101} (\bibinfo {year} {2009})}\BibitemShut {NoStop}%
\bibitem [{\citenamefont {Vuatelet}\ and\ \citenamefont
  {Ran{\c{c}}on}(2021)}]{MBDL_powerdecay_rancon}%
  \BibitemOpen
  \bibfield  {author} {\bibinfo {author} {\bibfnamefont {V.}~\bibnamefont
  {Vuatelet}}\ and\ \bibinfo {author} {\bibfnamefont {A.}~\bibnamefont
  {Ran{\c{c}}on}},\ }\href@noop {} {\bibinfo {title} {Effective thermalization
  of a many-body dynamically localized bose gas}} (\bibinfo {year} {2021}),\
  \Eprint {https://arxiv.org/abs/2103.14388} {arXiv:2103.14388
  [cond-mat.quant-gas]} \BibitemShut {NoStop}%
\bibitem [{\citenamefont {Rozenbaum}\ \emph {et~al.}(2017)\citenamefont
  {Rozenbaum}, \citenamefont {Ganeshan},\ and\ \citenamefont
  {Galitski}}]{lyapunovotocqkr_victor}%
  \BibitemOpen
  \bibfield  {author} {\bibinfo {author} {\bibfnamefont {E.~B.}\ \bibnamefont
  {Rozenbaum}}, \bibinfo {author} {\bibfnamefont {S.}~\bibnamefont
  {Ganeshan}},\ and\ \bibinfo {author} {\bibfnamefont {V.}~\bibnamefont
  {Galitski}},\ }\href {https://doi.org/10.1103/PhysRevLett.118.086801}
  {\bibfield  {journal} {\bibinfo  {journal} {Phys. Rev. Lett.}\ }\textbf
  {\bibinfo {volume} {118}},\ \bibinfo {pages} {086801} (\bibinfo {year}
  {2017})}\BibitemShut {NoStop}%
\bibitem [{\citenamefont {Gorin}\ \emph {et~al.}(2006)\citenamefont {Gorin},
  \citenamefont {Prosen}, \citenamefont {Seligman},\ and\ \citenamefont
  {Žnidarič}}]{loschmidt_review}%
  \BibitemOpen
  \bibfield  {author} {\bibinfo {author} {\bibfnamefont {T.}~\bibnamefont
  {Gorin}}, \bibinfo {author} {\bibfnamefont {T.}~\bibnamefont {Prosen}},
  \bibinfo {author} {\bibfnamefont {T.~H.}\ \bibnamefont {Seligman}},\ and\
  \bibinfo {author} {\bibfnamefont {M.}~\bibnamefont {Žnidarič}},\ }\href
  {https://doi.org/https://doi.org/10.1016/j.physrep.2006.09.003} {\bibfield
  {journal} {\bibinfo  {journal} {Phys. Rep.}\ }\textbf {\bibinfo {volume}
  {435}},\ \bibinfo {pages} {33} (\bibinfo {year} {2006})}\BibitemShut
  {NoStop}%
\bibitem [{\citenamefont {Ullah}\ and\ \citenamefont
  {Hoogerland}(2011)}]{loschmidtcooling_exp}%
  \BibitemOpen
  \bibfield  {author} {\bibinfo {author} {\bibfnamefont {A.}~\bibnamefont
  {Ullah}}\ and\ \bibinfo {author} {\bibfnamefont {M.~D.}\ \bibnamefont
  {Hoogerland}},\ }\href {https://doi.org/10.1103/PhysRevE.83.046218}
  {\bibfield  {journal} {\bibinfo  {journal} {Phys. Rev. E}\ }\textbf {\bibinfo
  {volume} {83}},\ \bibinfo {pages} {046218} (\bibinfo {year}
  {2011})}\BibitemShut {NoStop}%
\bibitem [{\citenamefont {Martin}\ \emph {et~al.}(2008)\citenamefont {Martin},
  \citenamefont {Georgeot},\ and\ \citenamefont
  {Shepelyansky}}]{loschmidtcooling_theory}%
  \BibitemOpen
  \bibfield  {author} {\bibinfo {author} {\bibfnamefont {J.}~\bibnamefont
  {Martin}}, \bibinfo {author} {\bibfnamefont {B.}~\bibnamefont {Georgeot}},\
  and\ \bibinfo {author} {\bibfnamefont {D.~L.}\ \bibnamefont {Shepelyansky}},\
  }\href {https://doi.org/10.1103/PhysRevLett.100.044106} {\bibfield  {journal}
  {\bibinfo  {journal} {Phys. Rev. Lett.}\ }\textbf {\bibinfo {volume} {100}},\
  \bibinfo {pages} {044106} (\bibinfo {year} {2008})}\BibitemShut {NoStop}%
\bibitem [{\citenamefont {Lepers}\ \emph {et~al.}(2008)\citenamefont {Lepers},
  \citenamefont {Zehnl\'e},\ and\ \citenamefont
  {Garreau}}]{positionspace_quantumresonance}%
  \BibitemOpen
  \bibfield  {author} {\bibinfo {author} {\bibfnamefont {M.}~\bibnamefont
  {Lepers}}, \bibinfo {author} {\bibfnamefont {V.}~\bibnamefont {Zehnl\'e}},\
  and\ \bibinfo {author} {\bibfnamefont {J.~C.}\ \bibnamefont {Garreau}},\
  }\href {https://doi.org/10.1103/PhysRevA.77.043628} {\bibfield  {journal}
  {\bibinfo  {journal} {Phys. Rev. A}\ }\textbf {\bibinfo {volume} {77}},\
  \bibinfo {pages} {043628} (\bibinfo {year} {2008})}\BibitemShut {NoStop}%
\bibitem [{\citenamefont {Swingle}\ \emph {et~al.}(2016)\citenamefont
  {Swingle}, \citenamefont {Bentsen}, \citenamefont {Schleier-Smith},\ and\
  \citenamefont {Hayden}}]{otoc2016PRA}%
  \BibitemOpen
  \bibfield  {author} {\bibinfo {author} {\bibfnamefont {B.}~\bibnamefont
  {Swingle}}, \bibinfo {author} {\bibfnamefont {G.}~\bibnamefont {Bentsen}},
  \bibinfo {author} {\bibfnamefont {M.}~\bibnamefont {Schleier-Smith}},\ and\
  \bibinfo {author} {\bibfnamefont {P.}~\bibnamefont {Hayden}},\ }\href
  {https://doi.org/10.1103/PhysRevA.94.040302} {\bibfield  {journal} {\bibinfo
  {journal} {Phys. Rev. A}\ }\textbf {\bibinfo {volume} {94}},\ \bibinfo
  {pages} {040302} (\bibinfo {year} {2016})}\BibitemShut {NoStop}%
\bibitem [{\citenamefont {Toh}\ \emph {et~al.}(2021)\citenamefont {Toh},
  \citenamefont {McCormick}, \citenamefont {Tang}, \citenamefont {Su},
  \citenamefont {Luo}, \citenamefont {Zhang},\ and\ \citenamefont
  {Gupta}}]{interactingqkr_deep}%
  \BibitemOpen
  \bibfield  {author} {\bibinfo {author} {\bibfnamefont {J.~H.~S.}\
  \bibnamefont {Toh}}, \bibinfo {author} {\bibfnamefont {K.~C.}\ \bibnamefont
  {McCormick}}, \bibinfo {author} {\bibfnamefont {X.}~\bibnamefont {Tang}},
  \bibinfo {author} {\bibfnamefont {Y.}~\bibnamefont {Su}}, \bibinfo {author}
  {\bibfnamefont {X.-W.}\ \bibnamefont {Luo}}, \bibinfo {author} {\bibfnamefont
  {C.}~\bibnamefont {Zhang}},\ and\ \bibinfo {author} {\bibfnamefont
  {S.}~\bibnamefont {Gupta}},\ }\href@noop {} {\bibinfo {title} {Observation of
  many-body dynamical delocalization in a kicked ultracold gas}} (\bibinfo
  {year} {2021}),\ \Eprint {https://arxiv.org/abs/2106.13773} {arXiv:2106.13773
  [cond-mat.quant-gas]} \BibitemShut {NoStop}%
\bibitem [{\citenamefont {Pollack}\ \emph {et~al.}(2009)\citenamefont
  {Pollack}, \citenamefont {Dries}, \citenamefont {Junker}, \citenamefont
  {Chen}, \citenamefont {Corcovilos},\ and\ \citenamefont
  {Hulet}}]{hulet_feshbach}%
  \BibitemOpen
  \bibfield  {author} {\bibinfo {author} {\bibfnamefont {S.~E.}\ \bibnamefont
  {Pollack}}, \bibinfo {author} {\bibfnamefont {D.}~\bibnamefont {Dries}},
  \bibinfo {author} {\bibfnamefont {M.}~\bibnamefont {Junker}}, \bibinfo
  {author} {\bibfnamefont {Y.~P.}\ \bibnamefont {Chen}}, \bibinfo {author}
  {\bibfnamefont {T.~A.}\ \bibnamefont {Corcovilos}},\ and\ \bibinfo {author}
  {\bibfnamefont {R.~G.}\ \bibnamefont {Hulet}},\ }\href
  {https://doi.org/10.1103/PhysRevLett.102.090402} {\bibfield  {journal}
  {\bibinfo  {journal} {Phys. Rev. Lett.}\ }\textbf {\bibinfo {volume} {102}},\
  \bibinfo {pages} {090402} (\bibinfo {year} {2009})}\BibitemShut {NoStop}%
\bibitem [{\citenamefont {Deuar}\ and\ \citenamefont
  {Drummond}(2007)}]{positivep-theory}%
  \BibitemOpen
  \bibfield  {author} {\bibinfo {author} {\bibfnamefont {P.}~\bibnamefont
  {Deuar}}\ and\ \bibinfo {author} {\bibfnamefont {P.~D.}\ \bibnamefont
  {Drummond}},\ }\href {https://doi.org/10.1103/PhysRevLett.98.120402}
  {\bibfield  {journal} {\bibinfo  {journal} {Phys. Rev. Lett.}\ }\textbf
  {\bibinfo {volume} {98}},\ \bibinfo {pages} {120402} (\bibinfo {year}
  {2007})}\BibitemShut {NoStop}%
\bibitem [{\citenamefont {Tenart}\ \emph {et~al.}(2020)\citenamefont {Tenart},
  \citenamefont {Carcy}, \citenamefont {Cayla}, \citenamefont {Bourdel},
  \citenamefont {Mancini},\ and\ \citenamefont {Cl\'ement}}]{2body_exp}%
  \BibitemOpen
  \bibfield  {author} {\bibinfo {author} {\bibfnamefont {A.}~\bibnamefont
  {Tenart}}, \bibinfo {author} {\bibfnamefont {C.}~\bibnamefont {Carcy}},
  \bibinfo {author} {\bibfnamefont {H.}~\bibnamefont {Cayla}}, \bibinfo
  {author} {\bibfnamefont {T.}~\bibnamefont {Bourdel}}, \bibinfo {author}
  {\bibfnamefont {M.}~\bibnamefont {Mancini}},\ and\ \bibinfo {author}
  {\bibfnamefont {D.}~\bibnamefont {Cl\'ement}},\ }\href
  {https://doi.org/10.1103/PhysRevResearch.2.013017} {\bibfield  {journal}
  {\bibinfo  {journal} {Phys. Rev. Res.}\ }\textbf {\bibinfo {volume} {2}},\
  \bibinfo {pages} {013017} (\bibinfo {year} {2020})}\BibitemShut {NoStop}%
\bibitem [{\citenamefont {Gadway}\ \emph {et~al.}(2009)\citenamefont {Gadway},
  \citenamefont {Pertot}, \citenamefont {Reimann}, \citenamefont {Cohen},\ and\
  \citenamefont {Schneble}}]{gadway_kd}%
  \BibitemOpen
  \bibfield  {author} {\bibinfo {author} {\bibfnamefont {B.}~\bibnamefont
  {Gadway}}, \bibinfo {author} {\bibfnamefont {D.}~\bibnamefont {Pertot}},
  \bibinfo {author} {\bibfnamefont {R.}~\bibnamefont {Reimann}}, \bibinfo
  {author} {\bibfnamefont {M.~G.}\ \bibnamefont {Cohen}},\ and\ \bibinfo
  {author} {\bibfnamefont {D.}~\bibnamefont {Schneble}},\ }\href@noop {}
  {\bibfield  {journal} {\bibinfo  {journal} {Opt. Express}\ }\textbf {\bibinfo
  {volume} {17}},\ \bibinfo {pages} {19173} (\bibinfo {year}
  {2009})}\BibitemShut {NoStop}%
\bibitem [{\citenamefont {Ovchinnikov}\ \emph {et~al.}(1999)\citenamefont
  {Ovchinnikov}, \citenamefont {M\"uller}, \citenamefont {Doery}, \citenamefont
  {Vredenbregt}, \citenamefont {Helmerson}, \citenamefont {Rolston},\ and\
  \citenamefont {Phillips}}]{phillips_kd}%
  \BibitemOpen
  \bibfield  {author} {\bibinfo {author} {\bibfnamefont {Y.~B.}\ \bibnamefont
  {Ovchinnikov}}, \bibinfo {author} {\bibfnamefont {J.~H.}\ \bibnamefont
  {M\"uller}}, \bibinfo {author} {\bibfnamefont {M.~R.}\ \bibnamefont {Doery}},
  \bibinfo {author} {\bibfnamefont {E.~J.~D.}\ \bibnamefont {Vredenbregt}},
  \bibinfo {author} {\bibfnamefont {K.}~\bibnamefont {Helmerson}}, \bibinfo
  {author} {\bibfnamefont {S.~L.}\ \bibnamefont {Rolston}},\ and\ \bibinfo
  {author} {\bibfnamefont {W.~D.}\ \bibnamefont {Phillips}},\ }\href
  {https://doi.org/10.1103/PhysRevLett.83.284} {\bibfield  {journal} {\bibinfo
  {journal} {Phys. Rev. Lett.}\ }\textbf {\bibinfo {volume} {83}},\ \bibinfo
  {pages} {284} (\bibinfo {year} {1999})}\BibitemShut {NoStop}%
\bibitem [{\citenamefont {Klappauf}\ \emph {et~al.}(1999)\citenamefont
  {Klappauf}, \citenamefont {Oskay}, \citenamefont {Steck},\ and\ \citenamefont
  {Raizen}}]{momentumboundary_raizen}%
  \BibitemOpen
  \bibfield  {author} {\bibinfo {author} {\bibfnamefont {B.}~\bibnamefont
  {Klappauf}}, \bibinfo {author} {\bibfnamefont {W.}~\bibnamefont {Oskay}},
  \bibinfo {author} {\bibfnamefont {D.}~\bibnamefont {Steck}},\ and\ \bibinfo
  {author} {\bibfnamefont {M.}~\bibnamefont {Raizen}},\ }\href@noop {}
  {\bibfield  {journal} {\bibinfo  {journal} {Physica D}\ }\textbf {\bibinfo
  {volume} {131}},\ \bibinfo {pages} {78} (\bibinfo {year} {1999})}\BibitemShut
  {NoStop}%
\bibitem [{\citenamefont {Fishman}\ \emph {et~al.}(2003)\citenamefont
  {Fishman}, \citenamefont {Guarneri},\ and\ \citenamefont
  {Rebuzzini}}]{qam_theory}%
  \BibitemOpen
  \bibfield  {author} {\bibinfo {author} {\bibfnamefont {S.}~\bibnamefont
  {Fishman}}, \bibinfo {author} {\bibfnamefont {I.}~\bibnamefont {Guarneri}},\
  and\ \bibinfo {author} {\bibfnamefont {L.}~\bibnamefont {Rebuzzini}},\
  }\href@noop {} {\bibfield  {journal} {\bibinfo  {journal} {J. Stat. Phys.}\
  }\textbf {\bibinfo {volume} {110}},\ \bibinfo {pages} {911} (\bibinfo {year}
  {2003})}\BibitemShut {NoStop}%
\bibitem [{\citenamefont {Dana}\ and\ \citenamefont
  {Dorofeev}(2006)}]{kickedparticle_theory}%
  \BibitemOpen
  \bibfield  {author} {\bibinfo {author} {\bibfnamefont {I.}~\bibnamefont
  {Dana}}\ and\ \bibinfo {author} {\bibfnamefont {D.~L.}\ \bibnamefont
  {Dorofeev}},\ }\href {https://doi.org/10.1103/PhysRevE.73.026206} {\bibfield
  {journal} {\bibinfo  {journal} {Phys. Rev. E}\ }\textbf {\bibinfo {volume}
  {73}},\ \bibinfo {pages} {026206} (\bibinfo {year} {2006})}\BibitemShut
  {NoStop}%
\bibitem [{\citenamefont {D'Alessio}\ and\ \citenamefont
  {Rigol}(2014)}]{floquetthermalization_rigol}%
  \BibitemOpen
  \bibfield  {author} {\bibinfo {author} {\bibfnamefont {L.}~\bibnamefont
  {D'Alessio}}\ and\ \bibinfo {author} {\bibfnamefont {M.}~\bibnamefont
  {Rigol}},\ }\href {https://doi.org/10.1103/PhysRevX.4.041048} {\bibfield
  {journal} {\bibinfo  {journal} {Phys. Rev. X}\ }\textbf {\bibinfo {volume}
  {4}},\ \bibinfo {pages} {041048} (\bibinfo {year} {2014})}\BibitemShut
  {NoStop}%
\bibitem [{\citenamefont {Boness}\ \emph {et~al.}(2006)\citenamefont {Boness},
  \citenamefont {Bose},\ and\ \citenamefont {Monteiro}}]{kickedxxz1}%
  \BibitemOpen
  \bibfield  {author} {\bibinfo {author} {\bibfnamefont {T.}~\bibnamefont
  {Boness}}, \bibinfo {author} {\bibfnamefont {S.}~\bibnamefont {Bose}},\ and\
  \bibinfo {author} {\bibfnamefont {T.~S.}\ \bibnamefont {Monteiro}},\ }\href
  {https://doi.org/10.1103/PhysRevLett.96.187201} {\bibfield  {journal}
  {\bibinfo  {journal} {Phys. Rev. Lett.}\ }\textbf {\bibinfo {volume} {96}},\
  \bibinfo {pages} {187201} (\bibinfo {year} {2006})}\BibitemShut {NoStop}%
\bibitem [{\citenamefont {Boness}\ \emph {et~al.}(2010)\citenamefont {Boness},
  \citenamefont {Kudo},\ and\ \citenamefont {Monteiro}}]{kickedxxz2}%
  \BibitemOpen
  \bibfield  {author} {\bibinfo {author} {\bibfnamefont {T.}~\bibnamefont
  {Boness}}, \bibinfo {author} {\bibfnamefont {K.}~\bibnamefont {Kudo}},\ and\
  \bibinfo {author} {\bibfnamefont {T.~S.}\ \bibnamefont {Monteiro}},\ }\href
  {https://doi.org/10.1103/PhysRevE.81.046201} {\bibfield  {journal} {\bibinfo
  {journal} {Phys. Rev. E}\ }\textbf {\bibinfo {volume} {81}},\ \bibinfo
  {pages} {046201} (\bibinfo {year} {2010})}\BibitemShut {NoStop}%
\bibitem [{\citenamefont {Rigol}\ \emph {et~al.}(2008)\citenamefont {Rigol},
  \citenamefont {Dunjko},\ and\ \citenamefont
  {Olshanii}}]{rigol2008thermalization}%
  \BibitemOpen
  \bibfield  {author} {\bibinfo {author} {\bibfnamefont {M.}~\bibnamefont
  {Rigol}}, \bibinfo {author} {\bibfnamefont {V.}~\bibnamefont {Dunjko}},\ and\
  \bibinfo {author} {\bibfnamefont {M.}~\bibnamefont {Olshanii}},\ }\href@noop
  {} {\bibfield  {journal} {\bibinfo  {journal} {Nature}\ }\textbf {\bibinfo
  {volume} {452}},\ \bibinfo {pages} {854} (\bibinfo {year}
  {2008})}\BibitemShut {NoStop}%
\bibitem [{\citenamefont {Monroe}\ \emph {et~al.}(2021)\citenamefont {Monroe},
  \citenamefont {Campbell}, \citenamefont {Duan}, \citenamefont {Gong},
  \citenamefont {Gorshkov}, \citenamefont {Hess}, \citenamefont {Islam},
  \citenamefont {Kim}, \citenamefont {Linke}, \citenamefont {Pagano},
  \citenamefont {Richerme}, \citenamefont {Senko},\ and\ \citenamefont
  {Yao}}]{ionsimulator_review}%
  \BibitemOpen
  \bibfield  {author} {\bibinfo {author} {\bibfnamefont {C.}~\bibnamefont
  {Monroe}}, \bibinfo {author} {\bibfnamefont {W.~C.}\ \bibnamefont
  {Campbell}}, \bibinfo {author} {\bibfnamefont {L.-M.}\ \bibnamefont {Duan}},
  \bibinfo {author} {\bibfnamefont {Z.-X.}\ \bibnamefont {Gong}}, \bibinfo
  {author} {\bibfnamefont {A.~V.}\ \bibnamefont {Gorshkov}}, \bibinfo {author}
  {\bibfnamefont {P.~W.}\ \bibnamefont {Hess}}, \bibinfo {author}
  {\bibfnamefont {R.}~\bibnamefont {Islam}}, \bibinfo {author} {\bibfnamefont
  {K.}~\bibnamefont {Kim}}, \bibinfo {author} {\bibfnamefont {N.~M.}\
  \bibnamefont {Linke}}, \bibinfo {author} {\bibfnamefont {G.}~\bibnamefont
  {Pagano}}, \bibinfo {author} {\bibfnamefont {P.}~\bibnamefont {Richerme}},
  \bibinfo {author} {\bibfnamefont {C.}~\bibnamefont {Senko}},\ and\ \bibinfo
  {author} {\bibfnamefont {N.~Y.}\ \bibnamefont {Yao}},\ }\href
  {https://doi.org/10.1103/RevModPhys.93.025001} {\bibfield  {journal}
  {\bibinfo  {journal} {Rev. Mod. Phys.}\ }\textbf {\bibinfo {volume} {93}},\
  \bibinfo {pages} {025001} (\bibinfo {year} {2021})}\BibitemShut {NoStop}%
\bibitem [{\citenamefont {Browaeys}\ and\ \citenamefont
  {Lahaye}(2020)}]{rydbergsimulator_review}%
  \BibitemOpen
  \bibfield  {author} {\bibinfo {author} {\bibfnamefont {A.}~\bibnamefont
  {Browaeys}}\ and\ \bibinfo {author} {\bibfnamefont {T.}~\bibnamefont
  {Lahaye}},\ }\href@noop {} {\bibfield  {journal} {\bibinfo  {journal} {Nat.
  Phys.}\ }\textbf {\bibinfo {volume} {16}},\ \bibinfo {pages} {132} (\bibinfo
  {year} {2020})}\BibitemShut {NoStop}%
\bibitem [{\citenamefont {Salath\'e}\ \emph {et~al.}(2015)\citenamefont
  {Salath\'e}, \citenamefont {Mondal}, \citenamefont {Oppliger}, \citenamefont
  {Heinsoo}, \citenamefont {Kurpiers}, \citenamefont
  {Poto\ifmmode~\check{c}\else \v{c}\fi{}nik}, \citenamefont {Mezzacapo},
  \citenamefont {Las~Heras}, \citenamefont {Lamata}, \citenamefont {Solano},
  \citenamefont {Filipp},\ and\ \citenamefont
  {Wallraff}}]{superconducting_spinmodelsim}%
  \BibitemOpen
  \bibfield  {author} {\bibinfo {author} {\bibfnamefont {Y.}~\bibnamefont
  {Salath\'e}}, \bibinfo {author} {\bibfnamefont {M.}~\bibnamefont {Mondal}},
  \bibinfo {author} {\bibfnamefont {M.}~\bibnamefont {Oppliger}}, \bibinfo
  {author} {\bibfnamefont {J.}~\bibnamefont {Heinsoo}}, \bibinfo {author}
  {\bibfnamefont {P.}~\bibnamefont {Kurpiers}}, \bibinfo {author}
  {\bibfnamefont {A.}~\bibnamefont {Poto\ifmmode~\check{c}\else
  \v{c}\fi{}nik}}, \bibinfo {author} {\bibfnamefont {A.}~\bibnamefont
  {Mezzacapo}}, \bibinfo {author} {\bibfnamefont {U.}~\bibnamefont
  {Las~Heras}}, \bibinfo {author} {\bibfnamefont {L.}~\bibnamefont {Lamata}},
  \bibinfo {author} {\bibfnamefont {E.}~\bibnamefont {Solano}}, \bibinfo
  {author} {\bibfnamefont {S.}~\bibnamefont {Filipp}},\ and\ \bibinfo {author}
  {\bibfnamefont {A.}~\bibnamefont {Wallraff}},\ }\href
  {https://doi.org/10.1103/PhysRevX.5.021027} {\bibfield  {journal} {\bibinfo
  {journal} {Phys. Rev. X}\ }\textbf {\bibinfo {volume} {5}},\ \bibinfo {pages}
  {021027} (\bibinfo {year} {2015})}\BibitemShut {NoStop}%
\bibitem [{\citenamefont {Sieberer}\ \emph {et~al.}(2019)\citenamefont
  {Sieberer}, \citenamefont {Olsacher}, \citenamefont {Elben}, \citenamefont
  {Heyl}, \citenamefont {Hauke}, \citenamefont {Haake},\ and\ \citenamefont
  {Zoller}}]{dqs_zoller}%
  \BibitemOpen
  \bibfield  {author} {\bibinfo {author} {\bibfnamefont {L.~M.}\ \bibnamefont
  {Sieberer}}, \bibinfo {author} {\bibfnamefont {T.}~\bibnamefont {Olsacher}},
  \bibinfo {author} {\bibfnamefont {A.}~\bibnamefont {Elben}}, \bibinfo
  {author} {\bibfnamefont {M.}~\bibnamefont {Heyl}}, \bibinfo {author}
  {\bibfnamefont {P.}~\bibnamefont {Hauke}}, \bibinfo {author} {\bibfnamefont
  {F.}~\bibnamefont {Haake}},\ and\ \bibinfo {author} {\bibfnamefont
  {P.}~\bibnamefont {Zoller}},\ }\href@noop {} {\bibfield  {journal} {\bibinfo
  {journal} {npj Quantum Inf.}\ }\textbf {\bibinfo {volume} {5}},\ \bibinfo
  {pages} {1} (\bibinfo {year} {2019})}\BibitemShut {NoStop}%
\bibitem [{\citenamefont {Goldman}\ and\ \citenamefont
  {Dalibard}(2014)}]{floquetgaugefields_goldman_dalibard}%
  \BibitemOpen
  \bibfield  {author} {\bibinfo {author} {\bibfnamefont {N.}~\bibnamefont
  {Goldman}}\ and\ \bibinfo {author} {\bibfnamefont {J.}~\bibnamefont
  {Dalibard}},\ }\href {https://doi.org/10.1103/PhysRevX.4.031027} {\bibfield
  {journal} {\bibinfo  {journal} {Phys. Rev. X}\ }\textbf {\bibinfo {volume}
  {4}},\ \bibinfo {pages} {031027} (\bibinfo {year} {2014})}\BibitemShut
  {NoStop}%
\end{thebibliography}%


\bibliographystyle{apsrev4-2}
\let\addcontentsline\oldaddcontentsline

\clearpage

\noindent
\textbf{Methods} \newline 
\noindent
\textbf{Experimental platform and sequence}. The experiments begin with a Bose-Einstein condensate (BEC) of around $10^5$ $^7$Li atoms in a far-detuned optical dipole trap with trapping frequencies $\omega_{x,z}/2\pi \approx 40$ Hz and $\omega_y/2\pi \approx 56$ Hz, where $z$ is the axis of the optical lattice, $y$ is the direction of gravity, and $x$ is the remaining orthogonal axis. The condensate is produced by optical evaporation at an $s$-wave scattering length of $a=240a_0$, set by an applied magnetic field in the vicinity of the broad Feshbach resonance at 737 Gauss~\cite{hulet_feshbach}. Immediately after evaporation, the fields are ramped to their desired value over 60-90 ms and maintained for the remainder of the experiment. The dipole trap is then extinguished and the BEC repeatedly subjected to a pulsed 1D optical lattice with lattice constant $d = 532$ nm, laser wave vector $k_{\mathrm{L}} = \pi/d$, and recoil energy $E_{\mathrm{R}} = \hbar^2 k_{\mathrm{L}}^2 / 2m$ with $m$ the mass of $^7$Li. The full dynamics are then well described by the second-quantized Hamiltonian
\begin{equation}
\begin{split}
    \mathcal{H} = \int\mathrm{d}^3 r \, \hat{\Psi}^{\dagger}(\vb{r},t) H(\vb{r},t) \hat{\Psi}(\vb{r},t) + \\ 
    \frac{g}{2} \int\mathrm{d}^3 r \, \hat{\Psi}^{\dagger}(\vb{r},t) \hat{\Psi}^{\dagger}(\vb{r},t) \hat{\Psi}(\vb{r},t) \hat{\Psi}(\vb{r},t) 
\end{split}
\label{eq:Hmanybody}
\end{equation}
\begin{align}
    H (\vb{r},t) = \frac{p^2}{2m} + \frac{V_0}{2} \cos(2 k_L z) I(x,y) \sum_n f_{\tau}(t-nT).
    \label{eq:Hsingleparticle}
\end{align}

The key kick parameters are the lattice depth $V_0$, effective pulse width $\tau$, and kick period $T$. $V_0$ is calibrated through a standard Kapitza-Dirac diffraction technique. $f_{\tau}(t)$ denotes a unit amplitude pulse function beginning at $t=0$ of width $\tau$. The experimental pulse is approximated by a piecewise function with a linear rise and fall of 200 ns duration before and after a plateau of variable hold time. For the experimental data in the main text with $\tau=300$ ns between the half-maximum points, this hold duration is 100 ns. The scattering length $a$ determines the two-body coupling coefficient $g=4 \pi \hbar^2 a/m$. Here $I(x,y)$ denotes the transverse intensity profile of the lattice beams normalized to unity maximum; this is approximately Gaussian $I(x,y) \approx e^{-2(x^2+y^2)/\sigma^2}$ with a measured 1/$e^2$ beam radius of $\sigma \approx 65$ $\upmu$m. The total duration of kicking is at most 1 ms for our longest experiments, significantly shorter than the 4 ms it takes the BEC to fall under the influence of gravity through the lattice beam waist. 

To measure the momentum distribution, we perform absorption imaging of the atoms after free expansion.   The time-of-flight (TOF) duration is 3.5 ms for the delocalization data and 2 ms for the Loschmidt data. Due to the low mass of $^7$Li and the breadth of the Feshbach resonance, coil inductance prevents sweeping the magnetic fields to the noninteracting regime for this expansion period. This means additional scattering occurs during expansion, which may lead to systematic errors in the measured quantities (see supplementary section \ref{sec:scatteringloschmidt}). For the energy, we are able to account for this scattering in our analysis due to the energy-conserving nature of the collisions. For metrics such as the IPR, this systematic is challenging to avoid. However by tracking the evolution of these observables as a function of kick number at a fixed TOF duration, we can largely attribute the qualitative observed dynamics to the evolution under the Hamiltonian (\ref{eq:Hmanybody}) as opposed to the expansion. At large $n$, the majority of scattering happens during the kicking duration so expansion effects become negligible.

\setlength{\parskip}{1em}
\noindent
\textbf{Delocalization data analysis.} This section discusses the analysis behind Fig.~\ref{fig:delocalization}. Because the momentum distributions of the interacting samples change significantly over the course of the delocalization experiments, the quantities shown in Fig.~\ref{fig:delocalization} are computed directly from raw or averaged images as opposed to fitting procedures. However, this can make measurable quantities such as energy sensitive to noise, especially near the edge of the camera sensor due to the quadratic weighting. To maximize the signal-to-noise ratio in our measurement, we analyze raw images using an adaptive region-of-interest (ROI). First, a single base ROI capturing all detectable atoms at all times is created for each interaction strength. The integrated density in this ROI is used to post-select images with total atom numbers falling within a $\pm 10$\% window of the mean, in order to reduce variations in the interaction energy, which depends directly on atom density. For these data we take 10 images at each kick number, of which typically 4-7 are discarded by this post-selection procedure. The ROI boundaries at each kick number are then determined by the points at which the cumulative summed distributions of the averaged image outward from a center point reach a threshold value. The thresholds are set empirically and the boundaries obtained by the following procedure. First we compute the transverse bound by integrating out the entire axial direction to get the overall transverse distribution, find the point it crosses an 85\% threshold and then expand the resulting boundary by a factor of 1.5 (1.2 in the delocalization data of supplementary section \ref{sec:T2.2data}) to ensure all atoms are captured. We then compute an axial boundary going point by point along the transverse direction; at each transverse point we integrate over 10 neighboring transverse pixel rows to get a ``local" axial distribution, find the point it crosses a 99.8\% threshold and expand by a factor 1.15. Finally we smooth each ROI boundary and take a moving average across different kick numbers (4 on each side). Crucially, we have confirmed that the qualitative observation of delocalization is not significantly altered from the simple case where we use just the initial single base ROI across all shots. However, the details of the trends should be more accurately captured by the adaptive procedure because the signal-to-noise ratio over the ROI is optimized at each kick number. All measurable quantities are then calculated from the imaged densities within this region.
\setlength{\parskip}{1pt}

Since we do not observe any substantial atom loss during the kicking duration, we treat the imaged atomic densities as normalized distributions. For Figs.~\ref{fig:delocalization}a-b, we compute the measured quantities from individual experimental runs and then average the results, with the reported error bar as the standard error of the mean. For Fig.~\ref{fig:delocalization}i, we instead compute the averaged distributions first before computing the deviation from exponential localization; the errorbars are computed from a Monte-Carlo simulation of the uncertainty in $k_{\mathrm{loc}}$ discussed later in this section. A smoothing filter is applied to the displayed densities in Figs.~\ref{fig:delocalization}c-h for visual clarity, but not in the subsequent calculation of the localization deviation in Fig.~\ref{fig:delocalization}i.

To measure the energy, we compute the post-expansion spatial variance of the distribution in both the kicking $z$ and transverse $x$ directions of the image. Assuming cylindrical symmetry, the kinetic energy is then calculated as $m \left( \langle z^2 \rangle + 2 \langle x^2 \rangle \right) / 2 t_{\mathrm{TOF}}^2$ with $t_{\mathrm{TOF}} \approx 3.5$ ms (see section 2.4 for a discussion of possible corrections to the conversion of position to energy). For an accurate measurement of the interacting samples, inclusion of the transverse energy is necessary to account for energy-conserving scattering processes that occur both during the kicking and TOF. In addition, the inhomogeneous intensity profile of the beam $I(x,y)$ leads to a transverse energy oscillation in all samples including the noninteracting ones (see supplementary section \ref{sec:transversebeam}). Since we are not interested in this effect, we remove it to leading order by subtracting off the noninteracting transverse energy from each trace, so that the noninteracting energy is purely the kinetic energy along the kicking direction. To compute the error bars on the interacting data, we add the error of the total interacting energy and noninteracting transverse energy in quadrature. The single-particle localization energy $E_{\mathrm{loc}}$ is estimated by averaging the noninteracting trace for $n \geq 100$, and the reported uncertainty is based on the standard deviation of those points. We note that this uncertainty is not only due to experimental imperfections, but also due to natural dynamical fluctuations, as evidenced by the results of noninteracting simulations like those shown in Fig.~\ref{fig:varytau}.

We compute an effective 1D momentum-space IPR by first integrating out the transverse dimension and then summing the squares of the subsequent normalized axial density. We confirmed that this qualitatively matches the result of directly integrating the squared 2D distribution while largely eliminating the beam-induced transverse oscillation. Specifically for computing this metric, we apply a smoothing filter to the normalized densities consistently across all 3 interaction strengths. This suppresses high-frequency background noise which sets a lower bound on the measurable IPR due to the squaring procedure. The measured values are compared to two predictions based on an exponentially localized distribution. The blue shaded region is obtained by numerically computing the IPR for the momentum space distribution $\exp(-\abs{k}/k_{\mathrm{loc}}) \sum_j \exp(-(k-2k_{\mathrm{L}} j)^2/w^2)$, which models a Gaussian comb with an exponential envelope. This is a reasonable expectation for a finite-size, localized noninteracting condensate occupying only discrete momentum modes. The width parameter $w$ is measured from fitting the $n=0$ noninteracting condensate and takes into account the momentum-space resolution of the TOF given the finite condensate spatial extent. The width of the region is based on Monte Carlo simulation of uncertainty in $k_{\mathrm{loc}}$, where the resulting distribution is fit to a Gaussian to extract the mean and standard deviation. The green shaded region is calculated analytically for a pure exponential distribution of infinite extent and is given by $1/4k_{\mathrm{loc}}$. Taking into account the finite width of the imaging region changes the distribution normalization and leads to the following correction factor $(1 - \exp(-2 k_0/k_{\mathrm{loc}})/(1 - \exp(-k_0/k_{\mathrm{loc}}))^2$; here $k_0 \approx 9.85k_L$ is the half-width of our images which yields a negligible correction factor of $\approx 1.006$. The width of the region is computed through linearized error propagation.

In Fig.~\ref{fig:delocalization}i, the plotted localization metric is $\int_{-k_0}^{k_0} \max[r(k)-1,0] \dd k / 2 k_0$. Here, $r(k) = \abs{\psi(k)}^2/\exp(-k/k_{\mathrm{loc}})$ is the ratio of the measured axial density denoted $\abs{\psi(k)}^2$ and an exponential localization envelope. Here the maximum of $\abs{\psi(k)}^2$ is set to unity. Taking the maximum of $r(k)-1$ and 0 ensures that the result is only sensitive to regions of the distribution which decay more slowly than exponentially. That is, it interprets 0 as ``at least exponentially'' localized with respect to a given localization length, and thus characterizes departures from a given dynamically localized state in the traditional sense of exponentially localized wavefunctions. We note however that the system remaining exponentially localized but with a larger localization length would result in a non-zero value for this metric, which motivates the direct inspection of the distributions in Figs.~\ref{fig:delocalization}f-h. The reported values and errorbars are extracted by propagating a Gaussian uncertainty in the measured $k_{\mathrm{loc}}$ through a Monte-Carlo simulation. We find that the resulting distributions interpolate between sharply peaked at 0 with a rapid fall-off when well-localized, to positively skewed with non-zero peak in the delocalized regime. We empirically find that a log-normal distribution fits the Monte Carlo result well, and we use this fit to extract the data reported in Fig.~\ref{fig:delocalization}i. In particular, the markers indicate the mean of the distribution and the errorbars represent the interquartile range containing the central 50\% of the distribution. Because of the skewness, we investigate the Monte Carlo simulated distributions in more detail in supplementary section \ref{sec:montecarlo}.

\setlength{\parskip}{1em}
\noindent
\textbf{Loschmidt experimental sequence and data analysis.} Here we discuss the methods and analysis used to produce Fig.~\ref{fig:loschmidt}. The Loschmidt experiments begin similarly to the previously described sequence; for an $N$ kick Loschmidt sequence, the BEC is first kicked $N/2$ times near quantum resonance at the parameters $V_0 \approx 50  E_{\mathrm{R}}$, $\tau = 300$ ns, and $T = 9.93$ $\upmu$s. For this data, we  adjusted the lattice depth $V_0$ for different interaction strengths to achieve the same amount of absorbed energy after the first $N/2$ kicks. This compensates for a decrease in energy absorption at the same lattice depth for higher interaction strengths, which we attribute to the increase of the Thomas-Fermi radius of the BEC relative to the lattice beam size. Neglecting this effect would artificially enhance the fidelity at very large interaction strengths due to a reduction in the effective stochasticity parameter $K$. We plot the zero mode fraction after the first $N/2$ kicks without time-reversal (denoted $F'$ Gauss) in Fig.~\ref{fig:Loschmidt_compare} to benchmark this kicking amplitude normalization procedure.
\setlength{\parskip}{1pt}

After the first $N/2$ kicks, we wait a half period $T/2$ to shift the wavefunction spatially by half a lattice spacing, causing the sign of subsequent kicks to be reversed. We then apply another sequence of $N/2$ kicks using the same lattice parameters to complete the echo sequence. The time series in Fig.~\ref{fig:loschmidt}c-e show absorption images averaged over 5 shots for each kick number $n$ in a $N=10$ experiment. Since we begin with a zero-momentum condensate mode, to measure the Loschmidt fidelity we simply need to count the fraction of atoms remaining in this mode. While atoms in other momentum modes coupled by the lattice are easily distinguished, atoms that have undergone scattering events into a smeared-out background distribution are not always well-separated. Thus, to extract the return fraction we fit the axial atomic distribution around the zero-momentum mode with a pair of Gaussians of varying width. The narrower Gaussian accounts for atoms remaining in the zero-momentum condensate after expansion, while the broader Gaussian measures the atoms that have been collisionally ejected from the condensate~\cite{positivep-theory,2body_exp}. In Fig.~\ref{fig:loschmidt}b, we show the fraction of atoms remaining in the narrow Gaussian and use this quantity as an estimate of the Loschmidt fidelity. Scattering during the expansion means that this necessarily underestimates the true fidelity, a possibility further addressed in supplementary section \ref{sec:scatteringloschmidt}.

\setlength{\parskip}{1em}
\noindent
\textbf{Noninteracting QKR numerics.} One-dimensional simulations of the noninteracting kicked rotor problem for comparison with experimental data are executed in two ways. We either perform a split-step Fourier integration of the QKR Hamiltonian (\ref{eq:Hsingleparticle}) (ignoring the transverse distribution $I(x,y)$) to model the finite-width pulse shapes, or iterate the QKR Floquet map described in the main text. The simulations are typically performed with periodic boundary conditions over a single lattice site (except when modeling the TOF readout; see supplementary section \ref{sec:positionspace}). We perform a Gaussian sampling of quasimomenta with standard deviation $\sim 0.1k_{\mathrm{L}}$, in rough accordance with the measured BEC temperature of around 10-15 nK. For simulation of the stochastic kicking protocol, we use the same techniques and additionally average over 100 different realizations of the fluctuations (note this is slightly different than in the experiment where a single kick period disorder realization is used).

\clearpage

\widetext
\begin{center}
    \textbf{\Large Supplementary Materials}
\end{center}
\setcounter{figure}{0}
\renewcommand{\figurename}{Fig.}
\makeatletter 
\renewcommand{\thefigure}{S\@arabic\c@figure}
\makeatother
\renewcommand{\thesection}{\arabic{section}}
\renewcommand{\thesubsection}{\thesection.\arabic{subsection}}
\renewcommand{\thesubsubsection}{\thesubsection.\arabic{subsubsection}}
\renewcommand{\theequation}{{S}\arabic{equation}}

\tableofcontents

\clearpage

%TC:ignore

%\section{Supplementary Text}
\section{Additional Delocalization Data}
\label{sec:T2.2data}

\begin{figure}[thb]
    \centering
    \includegraphics[scale=1]{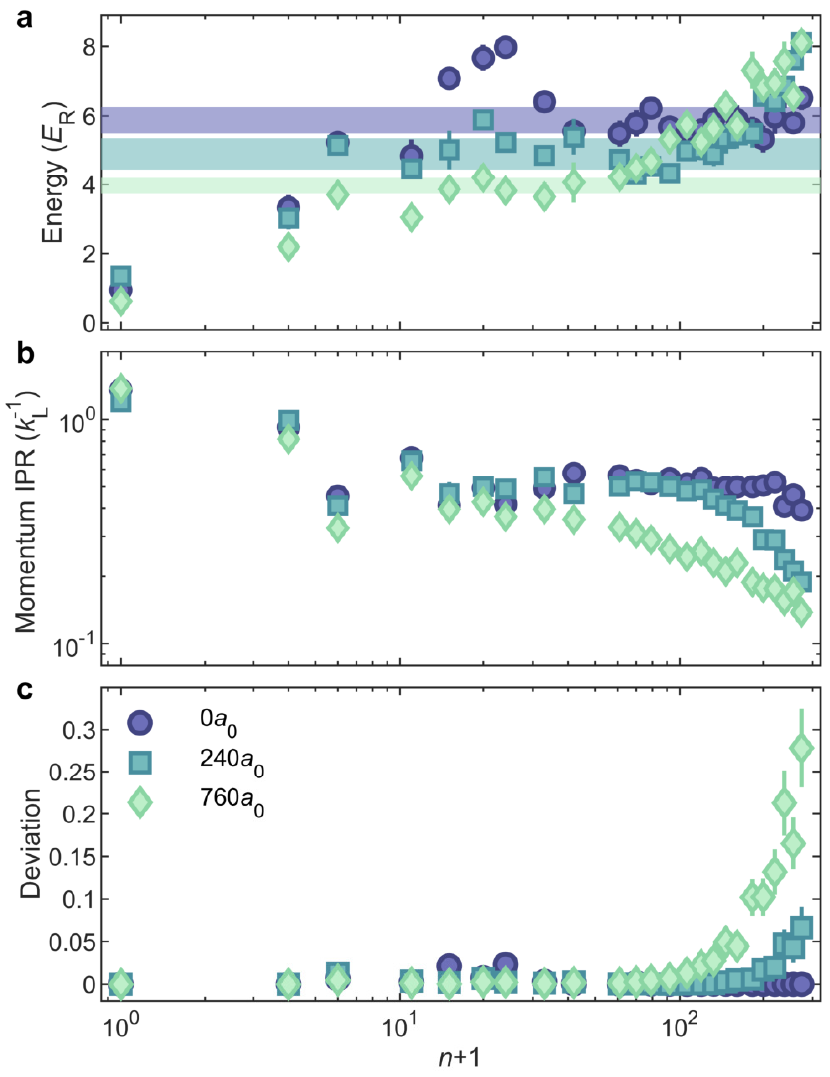}
    \caption{\textbf{Interaction-induced delocalization for a different set of kicking parameters.} The kick parameters are $V_0=70 \, E_{\mathrm{R}}$, $T=2.2$ $\upmu$s and $\tau=300$ ns ($K \approx 4.6$ and $\kbar \approx 2.8$). The (\textbf{a}) energy, (\textbf{b}) 1D momentum-space IPR and (\textbf{c}) deviation from exponential localization over time for varying scattering lengths. In a, the shaded regions indicate the extracted initial localization lengths for the three interaction strengths which we use for computing c.}
    \label{fig:t2.2}
\end{figure}

To supplement the dynamical delocalization signals shown in Fig.~\ref{fig:delocalization} and demonstrate that this is not a particularly fine-tuned phenomenon in the kicking parameter space, in Fig.~\ref{fig:t2.2} we show the same metrics for a larger kicking period $T=2.2$ $\upmu$s. The overall picture is unchanged, as the interacting samples show starkly different behavior from the noninteracting traces, departing from the localized value of each metric after a variable break time. Here the energy delocalization is obscured slightly as the different interaction strengths seem to initially localize to different energies. We attribute this partly to Thomas-Fermi expansion which reduces both the effective lattice depth experienced by the condensate and the initial kinetic energy of the sample, though we do not entirely rule out the possibility of different early-time prethermal behavior across interaction strengths. The correlation between localization length and quasimomentum spread is observed in noninteracting numerics. The different-colored shaded regions indicate our best estimates for the different localization energies at the 3 interaction strengths by computing the mean energy (and standard deviation) over windows of $n$ where the data are minimally changing. These values are used to compute the exponential localization deviation in Fig.~\ref{fig:t2.2}c. We do note a small trend visible at the end of the noninteracting traces; numerics suggest that this is consistent with variations in the localization length that occur over time for certain kicking parameters. 

\section{Monte Carlo distributions for exponential localization deviation}
\label{sec:montecarlo}
\begin{figure}
    \centering
    \includegraphics[scale=1]{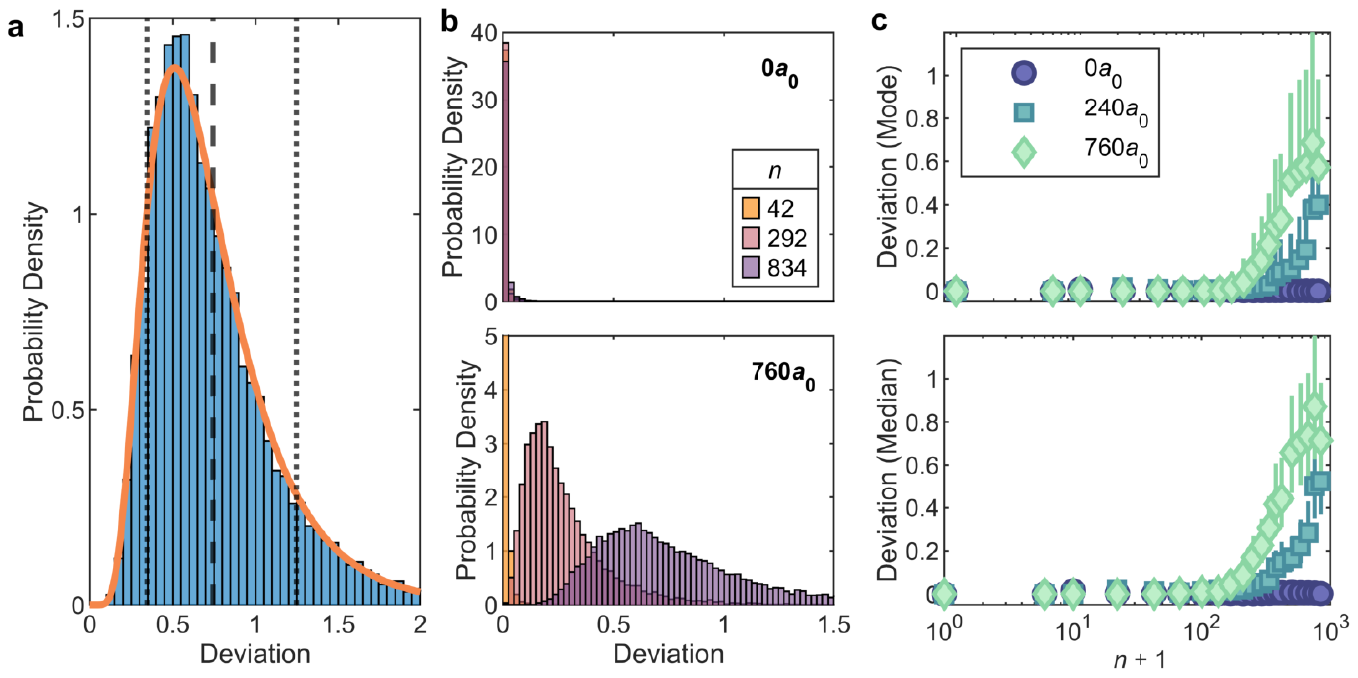}
    \caption{\textbf{Characterizing Monte Carlo distributions for deviation from exponential localization.} (\textbf{a}) Distribution for the 760$a_0$ data in Fig.~\ref{fig:delocalization} at $n=500$. Orange line indicates the fit to a log-normal distribution. Vertical dashed line indicates the mean, and the vertical dotted lines surrounding it indicate the interquartile range reported as the errorbars. (\textbf{b}) Evolution of the distribution over time for noninteracting and interacting samples. Note the difference in y-scale. The $n=42$ trace in the lower panel is cut-off vertically for visual clarity on the larger $n$ distributions. (\textbf{c}) Alternative characterizations of the exponential localization deviation in terms of the mode and median of the simulated distributions (indicated by the markers, the errorbars are left as the interquartile range), as opposed to the mean shown in the main text Fig.~\ref{fig:delocalization}i.}
    \label{fig:mc}
\end{figure}

In Fig.~\ref{fig:mc}, we show further details on the Monte Carlo simulated distributions for quantitatively characterizing the deviation from exponential localization in Figs.~\ref{fig:delocalization}i and \ref{fig:t2.2}c (in particular this data corresponds to Fig.~\ref{fig:delocalization}i). The distributions are generated by computing the defined deviation parameter for $10^4$ values of $k_{\mathrm{loc}}$ drawn from a Gaussian centered at 1.58$k_{\mathrm{L}}$ and with standard deviation 0.12$k_{\mathrm{L}}$. An example distribution for a sample which has delocalized is shown in Fig.~\ref{fig:mc}a, clearly showing the skewed probability densities we obtain from this procedure. The solid orange line indicates the log-normal distribution fit we use to extract parameters such as the mean and interquartile range of the distribution. We note that the use of a log-normal distribution here is only motivated empirically as a systematic method to determine such quantities.

In Fig.~\ref{fig:mc}b, we contrast how these simulated distributions evolve in time for localized noninteracting samples and delocalizing interacting ones. In the noninteracting case, the distributions are extremely sharply peaked at 0 and are relatively unchanging in-time, agreeing with the expectation of dynamical localization. In the latter, however, the distribution is only peaked at 0 for short times, a behavior indicative of the finite duration prethermal plateau we report, and gradually shifts away to non-zero values as the sample heats up. Importantly, at the later times the 760$a_0$ distribution has essentially vanishing probability density at 0 deviation, allowing us to confidently claim observation of departure from exponential localization. In Fig.~\ref{fig:mc}c, we confirm that the reported behavior of deviation over time in Fig.~\ref{fig:delocalization}I would not qualitatively change if we instead used the median or mode of the distribution instead of the mean.

\section{Systematics: Finite pulse width}
\label{sec:finitetau}

\begin{figure}[thb]
    \centering
    \includegraphics[scale = 1]{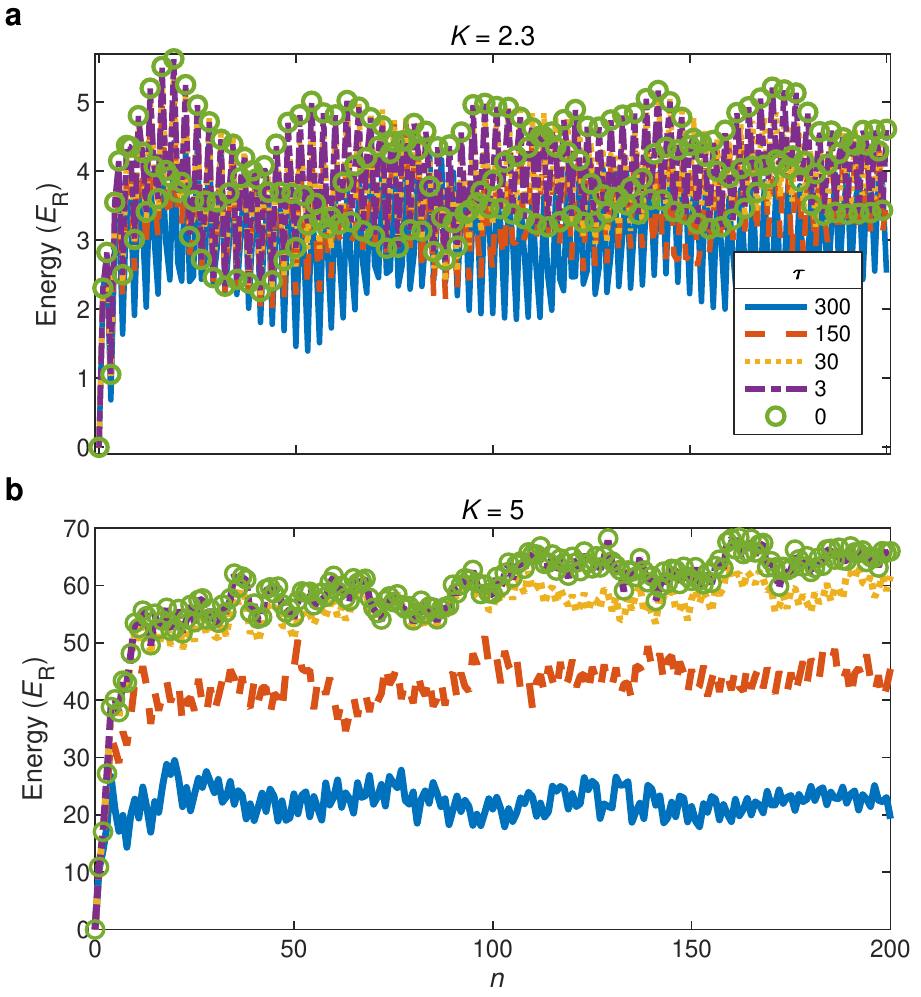}
    \caption{\textbf{Effects of finite pulse width on noninteracting QKR localization.} (\textbf{a}) Time evolution of energy for $K=2.3$. The $\tau=300$ ns trace is comparable to the data in the main text. (\textbf{b}) Equivalent simulation for $K=5$. The relative difference in localization energy between the achievable finite pulse durations in our experiment and the delta-kick limit becomes much more substantial at larger $K$.}
    \label{fig:varytau}
\end{figure}

%original para with incorrect RN regime
%The assumption corresponds to the Raman-Nath diffraction regime which is expressed by the condition $4 E_{\mathrm{R}} \tau/ \hbar \ll 1$. For the experiment with $\tau = 0.3$ $\upmu$s, this parameter is approximately 0.2. Excitation into higher momentum modes can also increase deviation from the delta-kick approximation.

%Correction to above using gadway expression: 
The delta-function kicking assumption in the theoretical QKR model is not perfectly realized in experiment owing to the finite atomic mass of $^7$Li. The assumption corresponds to the Raman-Nath diffraction regime which is approximately expressed by the condition $2\sqrt{V_0 E_{\mathrm{R}}} \tau/\hbar \ll 1$~\cite{gadway_kd}. For the experiment with $\tau = 0.3$ $\upmu$s and $V_0 = 64 E_{\mathrm{R}}$ as in Fig.~\ref{fig:delocalization}, this parameter is approximately 0.76 (there is an ambiguity of a factor $2\pi$ in defining the condition~\cite{phillips_kd}, which would reduce the parameter to 0.12). Either way, this suggests that the system is in between the Raman-Nath and Bragg diffraction regimes and thus finite-pulse-width effects require careful investigation. 

We numerically explore the effects of realistic pulse duration on single-particle QKR localization by comparing square pulse simulations of varying pulse width $\tau$ to the delta-kick Floquet map solution. To make this comparison, we keep the effective stochasticity parameter $K \sim V_0 \tau$ characterizing the kicking strength constant as we let $\tau \to 0$. Results for two values of $K$ are shown in Fig.~\ref{fig:varytau}. In general, we find that larger pulse duration tends to decrease the localization energy, which from a classical perspective corresponds to a particle traversing a significant part of the cosine potential during the kick and thus feeling a smaller effective impulse. This effect depends on the value of $K$ which determines the extent to which higher momentum modes are excited in the localized state. The many-body delocalization data in the main text were taken around $K \approx 2.3$. These simulations indicate that for this data there is a roughly 20\% decrease in the measured noninteracting localization energy with respect to the delta-kick limit. 

The finite pulse duration leads to an effective kicking strength which decays with increasing momentum, causing even the classical phase space to localize above a certain momentum~\cite{momentumboundary_raizen}. This is an important consideration in probing the destruction of the quantum dynamical localization which occurs in the classically chaotic regime. For our parameters, the estimate given in~\cite{momentumboundary_raizen} for the momentum boundary between classically chaotic and integrable regions due to pulse width is roughly $\pm 33.2 k_{\mathrm{L}}$, which is much larger than any excitation we observe in the data for Fig.~\ref{fig:delocalization} and \ref{fig:t2.2}. Thus we do not expect that the finite pulse duration qualitatively affects the observed delocalization dynamics.

We note that the reduction of absorbed energy by finite pulse width does play a practical role in determining which sets of system/kicking parameters are amenable to observation of interaction-induced delocalization. Because collisional processes are proportional to real-space density-density overlaps between different momentum modes, for a poor choice of kicking parameters a strong excitation of higher momentum modes in conjunction with real-space expansion (discussed in the next section) may rapidly dilute the system, and yield an effectively non-interacting sample before the interaction-induced delocalization break time.

\section{Systematics: Position space dynamics and TOF conversion} 
\label{sec:positionspace}

\begin{figure}[thb]
    \centering
    \includegraphics[scale = 1]{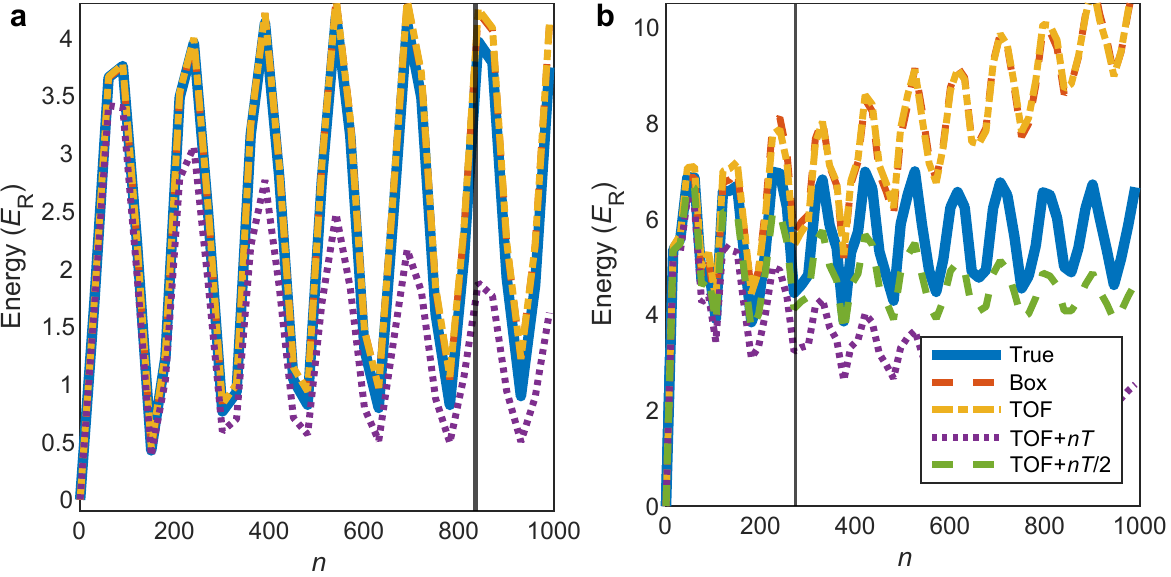}
    \caption{\textbf{Effects of position-space dynamics on extracting energy from TOF images.} Noninteracting numerical simulations comparing different methods for measuring energy in TOF and the true energy of the kicked rotor. (\textbf{a}) The kicking parameters are comparable to the ones used in the main text ($T=1.2 \, \upmu$s and $\tau=300\,$ns), but with lattice depth reduced to $V_0 = 38 E_{\mathrm{R}}$ to compensate for modeling only delta function kicking. We expect that the strong oscillations shown here are largely washed out in the experimental data by effective averaging over kicking strengths due to transverse extent of the condensate and fluctuations in the beam intensity from run-to-run. The traces here are sampled every 30 kicks for visual clarity, and the energy actually fluctuates at a much higher frequency. This simulation reveals that for the main data, the correct energy is straightforwardly extracted by simply using the TOF time to convert position to momentum. The vertical line indicates the maximum kick number reached in the data. (\textbf{b}) A similar simulation but with $T=2.2\,\upmu$s roughly corresponding to the supplementary delocalization data in Fig.~\ref{fig:t2.2} (with adjusted lattice depth $V_0 = 52 E_{\mathrm{R}}$). To illustrate the potential pitfalls of the readout method, here we model a TOF duration of only 2 ms, as opposed to the 3.5 ms used in the experiment. In this case, using only the TOF time in the velocity conversion leads to a false delocalization signal at late times. Instead, we show than an additional method using the TOF duration plus half the kicking time leads to the most faithful representation of the QKR energy for the longest time. We chose to use this conversion in the analysis of \ref{fig:t2.2}, though our simulations do not indicate a large difference between these two methods when modeling the full 3.5 ms TOF and restricting only to the max kick number indicated by the vertical line. Here the sampling is every 15 kicks for visual clarity.}
    \label{fig:tof}
\end{figure}

From a theoretical perspective, the difference between open and periodic boundary conditions in the single-particle QKR is resolved by Bloch's theorem~\cite{qam_theory,kickedparticle_theory}. Different quasimomenta evolve independently, manifesting different realizations of pseudo-randomness in the Anderson model mapping. The connection between the theoretical QKR and kicked quantum gas experiments is made by considering ensemble averages over quasimomenta. However, in practice, experimental readout of the kinetic energy even in the noninteracting case can be further complicated by spatial motion in a non-compact position variable during the course of the kicking. This effect was largely negligible for many previous QKR realizations using  heavy atomic species and/or short kicking durations, but in these experiments using light $^7$Li atoms and large kick numbers, careful consideration of the effect is required for an accurate energy measurement. Ideally, one would simply extend the TOF duration to suppress such effects, but technical limitations associated with the imaging procedure mean that this cannot be done indefinitely.

In interpreting the TOF absorption images as momentum space distributions, one must convert pixel position to velocity by dividing by an appropriate time. Without spatial motion, the correct time is trivially just the TOF duration. With spatial motion, a strict lower bound on the velocity conversion is set by the combined duration of the kicking and TOF which is equivalent to the assumption that each momentum mode propagates ballistically for the entire course of the experiment, ignoring the reshuffling of momentum modes by repeated kicking. To determine which conversion scheme leads to the most accurate energy measurement, we simulate the delta-function QKR model with an extended position space variable to model the TOF expansion explicitly. We are then able to compare the exact energy with various position-to-velocity conversions to determine the best metric. 

Simulations for different kicking parameters and TOF durations are shown in Fig.~\ref{fig:tof}a and b, revealing that in fact the simplest approach of using the TOF to convert position to momentum works well for the parameters of the experiments reported in the main text. For experiments at other parameter values, however, the appropriate conversion can change. In Fig.~\ref{fig:tof}a corresponding to the main data, we show that simply using the TOF as a conversion factor matches the true energy. For the simulations in \ref{fig:tof}b, however, adding in half of the kicking duration gives a substantially more accurate approximation of the true energy than the simple TOF conversion, which produces a false delocalization signal at longer times. We also examine ``box''-counting schemes where the image is instead binned into discrete modes which are multiples of $2k_{\mathrm{L}}$ momentum, though the added complexity of this scheme is not justified by the results.

\section{Systematics: Beam-induced transverse dynamics}
\label{sec:transversebeam}

\begin{figure}
    \centering
    \includegraphics[scale=1]{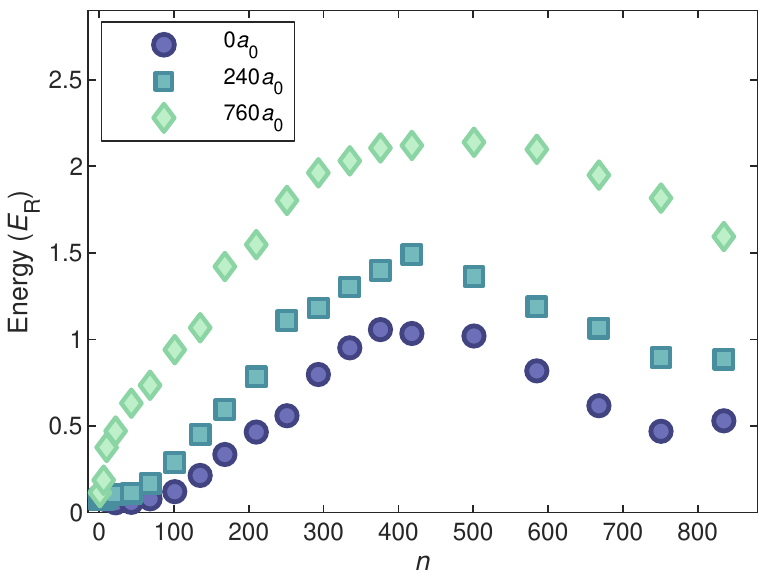}
    \caption{\textbf{Transverse dynamics in experimental quantum kicked rotors.} Measured transverse kinetic energy versus time for the main delocalization data in Fig.~\ref{fig:delocalization}. Each interaction strength undergoes an oscillation due to harmonic motion in the time-averaged lattice potential. The difference in evolution among different interaction strengths is due to scattering effects.}
    \label{fig:transverse}
\end{figure}

While the main concern of QKR experiments is with  momenta along the lattice direction, our experiments are three-dimensional and degrees of freedom transverse to the lattice beam cannot  in general be ignored especially in the presence of scattering. In Fig.~\ref{fig:transverse}, we explicitly show the measured \emph{transverse} kinetic energy for the main delocalization data in the text. Here we can see all three interaction strengths undergoing an oscillation in their transverse energy, which can be interpreted simply as harmonic motion in the time-averaged intensity distribution of the pulsed Gaussian lattice beam. The clear difference in the evolution between the different interaction strengths indicates the effects of 3D scattering for a system with uniform $I(x,y)$. As discussed in the Methods section 1.2, this motivates inclusion of the difference between the noninteracting and interacting transverse energy traces in the plotted energy of Fig.~\ref{fig:delocalization}. We have separately confirmed that ignoring the transverse dynamics altogether does not eliminate the observed delocalization signal.

\section{Systematics: Effects of scattering on measured Loschmidt fidelity}
\label{sec:scatteringloschmidt}
\begin{figure}
    \centering
    \includegraphics[scale=1]{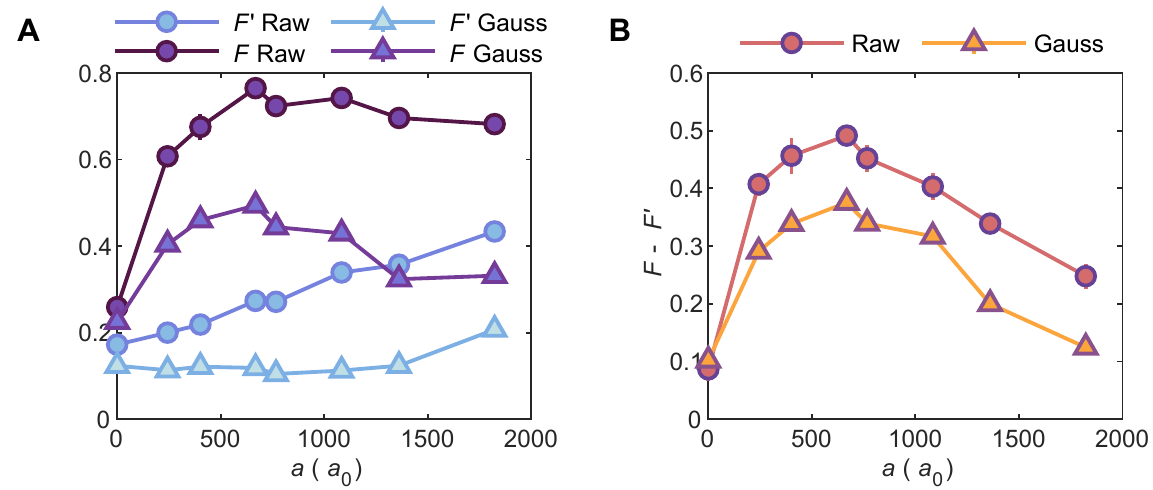}
    \caption{\textbf{Comparison of atom counting methods for Loschmidt fidelity.} (\textbf{A}) Zero mode fraction measured at the halfway point (\textit{F'}) and at the end (\textit{F}) of an 8-kick Loschmidt echo protocol for both raw and Gaussian counting methods. (\textbf{B}) The difference in zero mode fraction between \textit{F} and \textit{F'} in an 8-kick Loschmidt echo protocol for both methods.}
    \label{fig:Loschmidt_compare}
\end{figure}

Readout of the momentum distribution of interacting samples can be complicated by scattering during the TOF, and this particularly impacts the measurements of metrics such as IPR and return probability $F$, where the signatures of scattering events occurring at different stages of the experiment (i.e. during the lattice pulse trains versus during the time-of-flight) are not easily extracted from the resulting distribution. Here we examine how scattering affects the reported Loschmidt echo return probability shown in Fig.~\ref{fig:loschmidt}. In Fig.~\ref{fig:Loschmidt_compare} we present a comparison of two different methods for measuring the fidelity, which we argue should bound the true value and indicate the effect of this systematic. The \textit{Gaussian fitting} method was described in the methods section 1.3 and presented in Fig.~\ref{fig:loschmidt}. The \textit{raw counting} method computes the fidelity by integrating the raw distribution in a $\pm k_L$ width around the central mode. If all the scattering occurs prior to the TOF, then the \textit{Gaussian fitting} method is the appropriate counting procedure, as it discards all scattered atoms and only counts the remaining zero-order atoms. If, however, the majority of scattering occurs during the TOF, then this population should be included in the return probability, and so the \textit{raw counting} method would more accurately reflect the true fidelity. 

\begin{figure}
    \centering
    \includegraphics[scale=1]{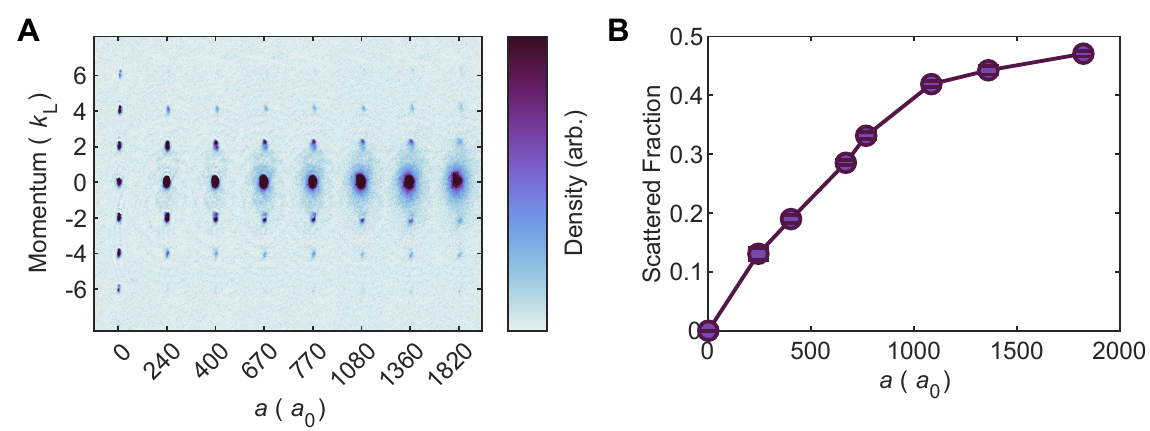}
    \caption{\textbf{Experimentally measured scattered fraction during Loschmidt experiments.} (\textbf{A}) TOF absorption images of a BEC after an $N = 8$ Loschmidt protocol at various scattering lengths. (\textbf{B}) Corresponding scattered fraction as computed using the \textit{Gaussian fitting} described in the methods section 1.3.} 
    \label{fig:Loschmidt_scatt}
\end{figure}

In Fig.~\ref{fig:Loschmidt_compare}A, we compare these two methods before (red) and after (blue) the application of the time-reversal kicks, as a function of scattering length. In both cases we find that the \textit{raw counting} method measures a higher fidelity than the \textit{Gaussian fitting} method due to accounting for the scattered population. As expected, the two converge in the noninteracting limit where the overall scattered population vanishes but diverge as the scattering length and consequently the scattered fraction increase (the dependence of scattered fraction on scattering length is plotted in Fig.~\ref{fig:Loschmidt_scatt}). This behavior of the \textit{Gaussian fitting} and \textit{raw counting} methods is  consistent with the limits of validity expected for each, and supports the claim that the two methods bound the systematic measurement error in counting the zero-order population that results from scattering during the TOF.

Having established approximate bounds for the true fidelity as a function of scattering length, we further remark that both methods produce a time-reversed fidelity which exhibits a crossover in behavior as a function of scattering length. The fidelity as a function of $a$ grows with weak interactions, and it saturates, or even decreases, as a function of $a$ with stronger interactions. We note also that imperfect calibration of the effective kicking strength could give rise to errors in the measured return fidelity. In Fig.~\ref{fig:Loschmidt_compare}B, we attempt to account for this effect across the two methods by considering the difference in fidelity before and after the set of time-reversal kicks. This produces two curves of similar functional form, further supporting our observation of a crossover into interaction-induced irreversibility.

\section{Mapping the QKR to a kicked spin chain model}
\label{sec:XXZmapping}

\begin{figure*}
    \centering
    \renewcommand{\figurename}{Fig.}
    \includegraphics[scale=1]{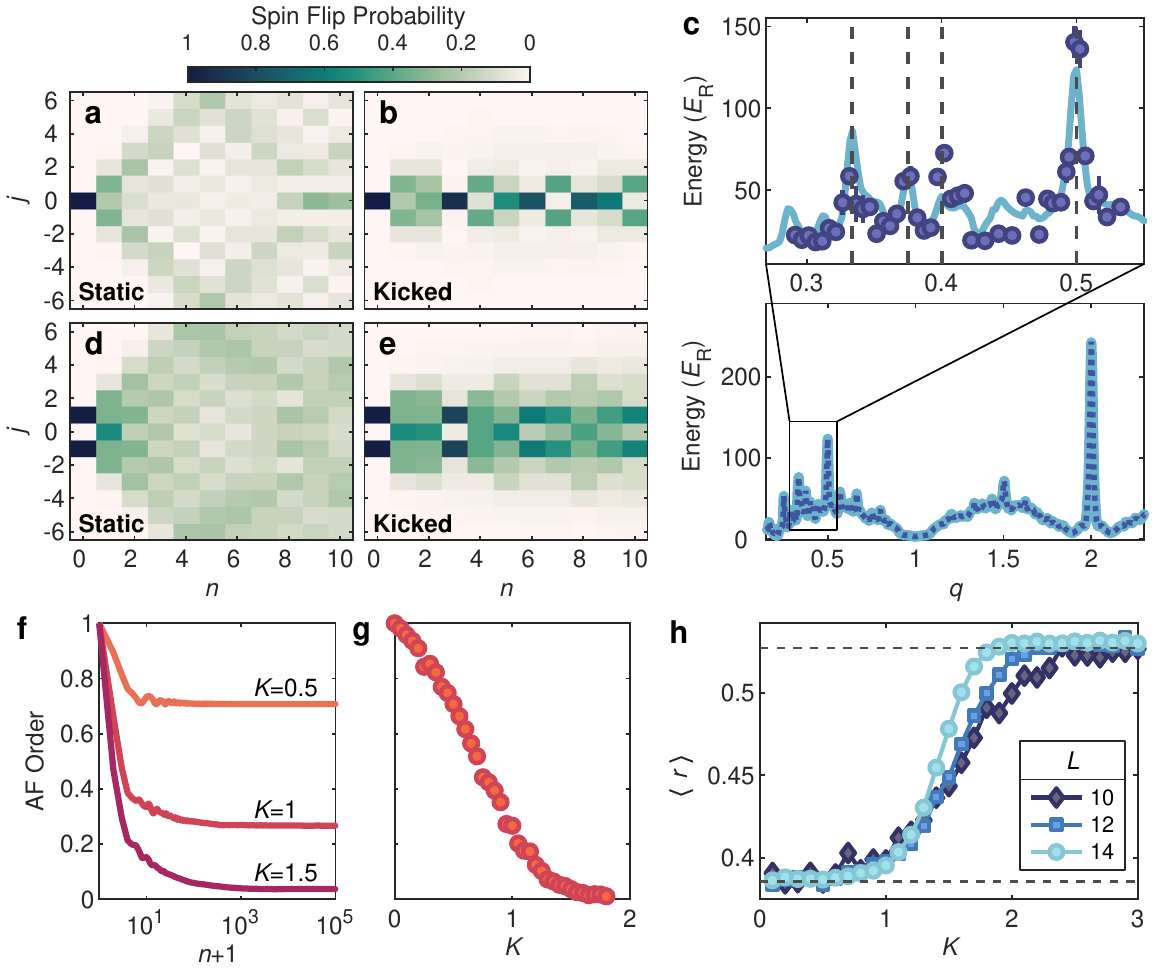}
    \caption{\textbf{Kicked-rotor dynamics and many-body dynamical localization in kicked spin chains.} (\textbf{a-b}) Calculated single-spin-flip evolution for static and kicked spin-1/2 XXX chains of length $L=13$ for $K=2.4$, $\kbar=1.5$ and $\beta=0$. (\textbf{c}) Spin chain quantum resonance spectroscopy (solid) demonstrating equivalence to QKR (dashed) in the single spin-flip sector with $K=3\kbar$, $L=51$ and averaging over a Gaussian ensemble of $\beta$. Energy is after 10 kicks. Zoomed-in panel compares experimental measurements on the atomic quantum kicked rotor to predictions of spin chain numerics. Dashed vertical lines indicate $q=$ 1/3, 3/8, 2/5, and 1/2. (\textbf{d-e}) The same as a-b but with 2 spin flips. (\textbf{f}) Time-averaged staggered magnetization versus kick number, starting from an initial N\'eel state with $\kbar=1$, $L=12$, $\beta=0.1$ and varying $K$. (\textbf{g}) Staggered magnetization in the infinite-time limit versus $K$, for the same parameters as panel f. (\textbf{h}) Gap-ratio statistic of the $M=0$ sector at $\kbar=1$ for varying $L$ averaged over 100 values of $\beta$. Dashed lines indicate predictions of the Poisson ($\langle r \rangle \approx 0.386$) and circular-orthogonal ensemble ($\langle r \rangle \approx 0.527$)~\cite{floquetthermalization_rigol}.}
    \label{fig:xxz}
\end{figure*}

The interplay between dynamical localization and quantum chaos is a topic of broad current interest, relevant in contexts well beyond the experimental model of the quantum kicked rotor which we explore here. To highlight this breadth, we describe and quantitatively explore a mapping from the QKR to a kicked Heisenberg spin chain, which can be used as a basis for generalization and further exploration. Specifically, we investigate the Floquet map $U_{\mathrm{kickedXXZ}} = e^{-i \kbar H_{\mathrm{quad}}/4} e^{-i K H_{\mathrm{XXZ}}/4\kbar}$, with $H_{\mathrm{quad}} = \sum_j (j+\beta)^2 \sigma_j^z$ a quadratic field and $H_{\mathrm{XXZ}} = \sum_j (\sigma_j^x \sigma_{j+1}^x + \sigma_j^y \sigma_{j+1}^y + \Delta \sigma_j^z \sigma_{j+1}^z)$ an XXZ Hamiltonian with $\sigma^{x,y,z}$ the Pauli matrices and $\Delta$ the anisotropy parameter. $\beta$ represents a field-center offset which serves as an analogue to quasimomentum in the standard QKR problem. Total magnetization $M=\langle \sum_j \sigma_j^z \rangle$ is a conserved quantity. This model is known to correspond to the QKR in single and few-body regimes~\cite{kickedxxz1,kickedxxz2}; for a single particle or spin flip the correspondence is essentially exact up to finite size effects. We perform exact diagonalization of $U_{\mathrm{kickedXXZ}}$ on systems of length up to $L=14$. We center the chains about $j=0$ (i.e. $j \in [1-L,L-1]/2$, integer for $L$ odd and half-integer for $L$ even) and consider open boundary conditions. To first illustrate the mapping, single-particle dynamical localization and quantum resonance in the spin model are demonstrated in Fig.~\ref{fig:xxz}a-c by considering the single spin-flip sector $M=-L+2$. In the zoom-in of Fig.~\ref{fig:xxz}c, we verify the QKR correspondence experimentally by comparing the kicked XXX ($\Delta=1$) numerics to our experimental observation of fractional $q$ quantum resonances in the QKR, and find excellent agreement. 

Given that the Heisenberg chain in a random magnetic field is a prototype model for traditional many-body localization (MBL), $U_{\mathrm{kickedXXZ}}$ appears well suited to address the question of whether the emergent Floquet pseudo-randomness in a disorder-free kicked system is sufficient to reproduce MBL phenomenology. To numerically probe many-body dynamical localization in the kicked spin model, we study the evolution of an initial N\'eel state ordering $\ket{\uparrow \downarrow \uparrow \downarrow \hdots}$ in the $M=0$ sector with multiple spin-flips, and observe long-lived persistence of ordering at finite $K$ values (Fig.~\ref{fig:xxz}f). As shown in Fig.~\ref{fig:xxz}g, this persistence lasts in the infinite time limit. These conclusions are obtained by computing the the time average of the staggered magnetization $O = \sum_j (-1)^j \sigma_j^z /L$ after $n$ cycles, given by $\langle O \rangle_n = \left( \sum_{i=1}^n \bra{\psi} U_{\mathrm{kickedXXZ}}^{\dagger i} O U_{\mathrm{kickedXXZ}}^i \ket{\psi}\right)/n$. The infinite-time limit is then calculated via the Floquet diagonal ensemble as $\lim_{n \to \infty} \langle O \rangle_n = \sum_{\alpha} \abs{c_{\alpha}}^2 \bra{\psi_{\alpha}} O \ket{\psi_{\alpha}}$. Here $c_{\alpha} = \bra{\psi_{\alpha}}\ket{\psi}$ are the coefficients of the initial state $\ket{\psi}$ in the basis of the many-body Floquet states $\ket{\psi_{\alpha}}$~\cite{rigol2008thermalization}. 

\begin{figure}
    \centering
    \includegraphics[scale=1]{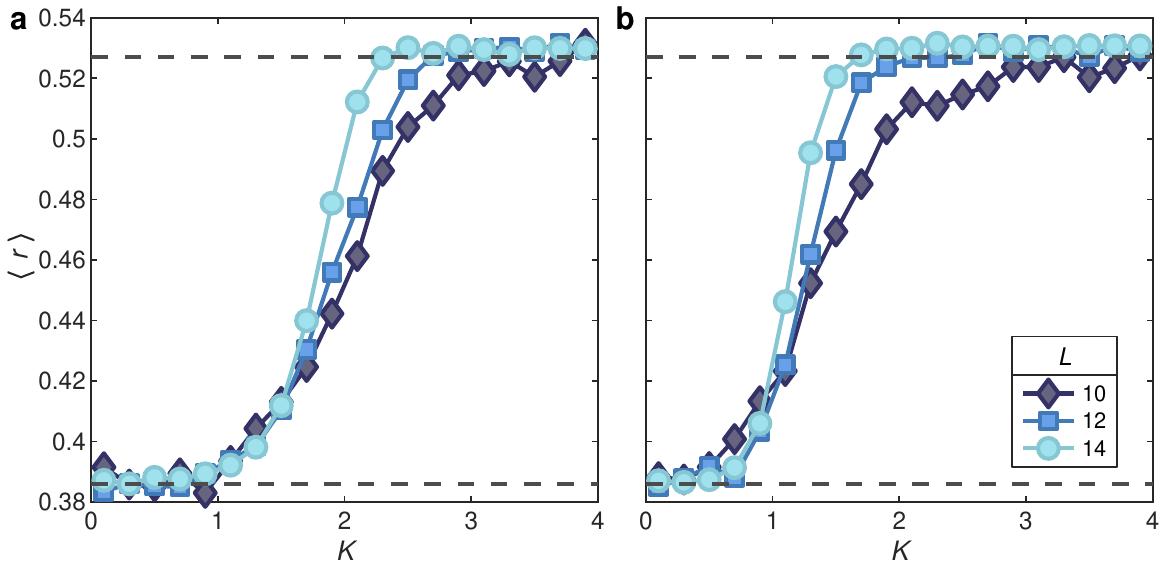}
    \caption{\textbf{Transition from MBDL to ergodicity for kicked XXZ model.} (\textbf{a}) Gap-ratio statistic for $\kbar=1$, $\Delta=0.5$ and averaging over a Gaussian ensemble of $\beta$ with standard deviation 0.1 for varying $K$ and $L$. (\textbf{b}) The same parameters except $\Delta=2$. Dashed horizontal lines indicate the same Poisson and circular-orthogonal ensemble predictions of Fig.~\ref{fig:xxz}h.}
    \label{fig:suppgapratio}
\end{figure}

A transition from many-body dynamical localization (MBDL) to ergodicity at larger interaction strengths is indicated both by the N\'eel state persistence (Fig.~\ref{fig:xxz}f-g) and by the Floquet level-spacing gap ratio parameter $\langle r \rangle$~\cite{floquetthermalization_rigol,supp} (Fig.~\ref{fig:xxz}h). The gap ratio is defined as $r_{\alpha} = \mathrm{min}\left(\delta_{\alpha},\delta_{\alpha+1}\right)/\mathrm{max}\left(\delta_{\alpha},\delta_{\alpha+1}\right)$, where $\delta_{\alpha} = \epsilon_{\alpha+1}-\epsilon_{\alpha}$ is the gap between consecutive quasi-energies $\epsilon_{\alpha}$, which we order on the interval $[-\pi,\pi]$. To compute $\langle r \rangle$, we average $r_{\alpha}$ over the $M=0$ sector as well as for 100 values of $\beta$ drawn from a normal distribution of standard deviation 0.1. We interpret the transition in $\langle r \rangle$ from the Poisson prediction to that of the circular-orthogonal ensemble as indication of distinct parameter regimes of MBDL and chaos in the many-body Floquet system. Similar results are found away from the isotropic $\Delta=1$ point, as shown in Fig.~\ref{fig:suppgapratio}. In the limit $\Delta=0$, we checked that the statistics are Poisson for all values of $K$ shown here. While extending to larger system sizes to extrapolate toward the thermodynamic limit remains an important task, these numerical results signal a true many-body dynamically localized state in kicked spin-chains with no disorder. This indicates a promising path towards experimentally observing MBDL in current-day quantum simulator platforms~\cite{ionsimulator_review,rydbergsimulator_review,superconducting_spinmodelsim} and highlights connections between paradigmatic kicked spin models and the quantum kicked rotor.

For the single-particle QKR, it is standard to consider periodic pulsing of the spatial potential separated by intervals of free kinetic energy evolution. In the spin-chain mapping, this corresponds to pulsed spin-exchange interactions and free evolution in a quadratic magnetic field. While here we used the exact mapping of the QKR parameters $K$ and $\kbar$ in $U$ to the spin model $U_{\mathrm{kickedXXZ}}$, physically we took the interpretation of an interacting XXZ Hamiltonian with a pulsed quadratic magnetic field, which we expect will be the most natural implementation in analog quantum platforms (the inherent Trotterization of a kicking Hamiltonian may also naturally lend itself to digital quantum simulation~\cite{dqs_zoller}). We remark that such a distinction is only manifest at the level of micromotion~\cite{floquetgaugefields_goldman_dalibard}. For the stroboscopic dynamics and Floquet level statistics analyzed in Fig. \ref{fig:xxz}, the problems are identical and simply require a relabeling of parameters.

%\subsection{MBDL for anisotropic kicked XXZ chains}

%Here we show that the isotropic $\Delta=1$ case considered in Fig.~\ref{fig:xxz} is not a fine-tuned point for realizing MBDL in the kicked XXZ model. In Fig.~\ref{fig:suppgapratio}A and B, we show the gap-ratio parameter computed for $\Delta = 0.5$ and $\Delta = 2$ respectively. In both cases we observe a transition from MBDL to chaos as the interaction strength $K$ is increased, which is qualitatively similar to that of the $\Delta=1$ case. We find that the region of $K$ for which the MBDL phase exists shrinks as the value of $\Delta$ is increased. For $\Delta=0$ (noninteracting), we find that $\langle r \rangle$ agrees with the Poisson prediction for all values of $K$ shown here.

%TC:endignore

\end{document}